\documentclass{aa}  
\usepackage{graphicx}
\usepackage[varg]{txfonts}
\usepackage{subfig}
\usepackage{newtxtext,newtxmath}
\usepackage[labelfont=bf]{caption}

\DeclareMathOperator{\sech}{sech}
\numberwithin{equation}{section} 
\newcommand{\kms}{~km~s$^{-1}$}

\usepackage[]{hyperref}
\hypersetup{colorlinks=true,linkcolor=red,citecolor=blue,urlcolor=magenta}

\defcitealias{2019Bacchini}{B19a}
\defcitealias{2019Bacchini_b}{B19b}

\begin{document}

   \title{The volumetric star formation law for nearby galaxies}
   \subtitle{Extension to dwarf galaxies and low-density regions}
    
   \authorrunning{C. Bacchini et al.}
   \author{Cecilia Bacchini\inst{1,2,3},
          Filippo Fraternali\inst{1},
           Gabriele Pezzulli\inst{4}, \and
           Antonino Marasco\inst{5}      
          }

   \institute{Kapteyn Astronomical Institute, University of Groningen, Landleven 12, 9747 AD Groningen, The Netherlands\\
   \email{c.bacchini@rug.nl}
   \and
   Dipartimento di Fisica e Astronomia, Universit\`{a} di Bologna, via Gobetti 93/2, I-40129, Bologna, Italy\\
   \email{cecilia.bacchini@unibo.it}
   \and
   INAF - Osservatorio di Astrofisica e Scienza dello Spazio di Bologna, via Gobetti 93/3, I-40129, Bologna, Italy
    \and
    Department of Physics, ETH Zurich, Wolfgang-Pauli-Strasse 27, 8093 Zurich, Switzerland
    \and
    INAF - Osservatorio Astrofisco di Arcetri, largo E. Fermi 5, 50127 Firenze, Italy
   }

   \date{Received 17 July 2020 / Accepted 15 October 2020.}

 
\abstract{In the last decades, much effort has been put into finding the star formation law which could unequivocally link the gas and the star formation rate (SFR) densities measured on sub-kiloparsec scale in star-forming galaxies. 
The conventional approach of using the observed surface densities to infer star formation laws has however revealed a major and well-known issue, as such relations are valid for the high-density regions of galaxies but break down in low-density and HI-dominated environments. 
Recently, an empirical correlation between the total gas (HI+H$_2$) and the star formation rate (SFR) volume densities was obtained for a sample of nearby disc galaxies and for the Milky Way. 
This volumetric star formation (VSF) law is a single power-law with no break and a smaller intrinsic scatter with respect to the star formation laws based on the surface density. 
In this work, we explore the VSF law in the regime of dwarf galaxies in order to test its validity in HI-dominated, low-density, and low-metallicity environments. 
In addition, we assess this relation in the outskirts of spiral galaxies, which are low-density and HI-dominated regions similar to dwarf galaxies. 
Remarkably, we find the VSF law, namely $\rho_\mathrm{SFR} \propto \rho_\mathrm{gas}^\alpha$ with $\alpha \approx 2$, is valid for both these regimes. 
This result indicates that the VSF law, which holds unbroken for a wide range of gas ($\approx 3$~dex) and SFR ($\approx 6$~dex) volume densities, is the empirical relation with the smallest intrinsic scatter and is likely more fundamental than surface-based star formation laws.}


   \maketitle

\section{Introduction}\label{sec:intro}
The quest for a fundamental law of star formation on kiloparsec scale has been ongoing since sixty years ago, when \cite{1959Schmidt} proposed for the first time a power-law relation linking the star formation rate (SFR) and the atomic gas volume densities. 
The massive amount of literature on this topic reflects the importance of the star formation law in the context of galaxy formation and evolution. 
Indeed, some form of this relation can give useful constraints on the processes that regulate the conversion of gas into stars and is often implemented at sub-galactic scales, both in idealised models and cosmological simulations of galaxies. 
Thirty years after Schmidt's pioneering work, a turning point in the study of star formation laws was reached with the observational determination of the so-called Schmidt-Kennicutt (SK) law linking the gas (HI+H$_2$) and the SFR surface densities, which reads $\Sigma_\mathrm{SFR} \propto \Sigma_\mathrm{gas}^{N}$ with $N \approx 1.4$ \citep{1989Kennicutt,1998Kennicutt}. 
Until today, versions of the star formation law involving surface densities have been widely investigated and different formulations have been proposed. 
However, the validity of these relations seem to be contingent upon interstellar medium (ISM) conditions and galaxy type. 
In particular, many authors have reported that the star formation law derived for spiral galaxies does not hold in dwarf galaxies, suggesting the existence of different regimes of star formation for different galaxy types. 
We can characterise different surface-based star formation laws according to the quantities involved and the regime of validity. 

The first group of empirical relations involves the gas and the SFR surface densities averaged over the star-forming disc of a galaxy, which are connected by the so-called ``global'' SK law with $N \approx 1.4$. 
This relation was originally proposed by \cite{1998Kennicutt} based on a sample of normal spiral galaxies and infrared luminous starburts. 
While the Milky Way (MW) seems to follow this global SK law \citep[e.g.][]{2012KennicuttEvans}, dwarf and low-surface brightness galaxies are generally discordant, as their SFR surface densities tend to be lower than the values expected from the SK relation 
\citep[][but see \citealt{2016Teich} for a different conclusion]{2009Wyder,2013Gatto,2015Roychowdhury,2017Roychowdhury,2019delosReyesKennicutt}. 

The second category of star formation laws is based on the local surface densities (i.e. azimuthally averaged radial profiles, surface densities measured in kiloparsec or sub-kiloparsec regions) and was derived for spatially resolved galaxies \citep[e.g.][]{1989Kennicutt,2001MartinKennicutt}. 
Several authors found that even this ``local'' SK relation breaks down in low-density ($\Sigma_\mathrm{gas} \lesssim 10 \, \mathrm{M_\odot pc^{-2}}$) and HI-dominated environments, such as the outskirts of nearby spiral galaxies and of the MW (e.g. \citealt{2002Wong,2003Boissier,2006Misiriotis,2007Kennicutt,2008Bigiel}; \citealt{2019Bacchini}, hereafter \citetalias{2019Bacchini}; \citealt{2019Bacchini_b}, hereafter \citetalias{2019Bacchini_b}), low surface brightness galaxies \citep[e.g.][]{2008Boissier,2009Wyder}, and dwarf galaxies \citep[e.g.][]{2011Bolatto,2015Roychowdhury}. 
In these regions, the SFR surface density is very modest if compared to the values predicted by the local SK law and its correlation with the HI surface density is extremely weak or absent (e.g. \citealt{1998Ferguson,2010Bigiel,2011Bolatto,2011Schruba,2016Yim}; \citetalias{2019Bacchini,2019Bacchini_b}). 

The third group of star formation laws is based on the molecular gas surface density, namely $\Sigma_\mathrm{SFR} \propto \Sigma_\mathrm{H_2}^{N}$ with $N \approx 0.6-1.4$ \citep[e.g.][]{2002Wong,2004Heyer,2007Kennicutt,2008Bigiel,2012Marasco,2013Leroy,2016Watson,2019Bicalho,2019JimenezDonaire,2019delosReyesKennicutt,2020Kumari}. 
In particular, the emission lines of carbon-monoxide (CO) are typically used to trace the molecular gas distribution and, in order to derive the 
mass of the total molecular gas from the CO luminosity, it is necessary to estimate the CO-to-H$_2$ conversion factor ($\alpha_\mathrm{CO}$). Unfortunately, dwarf galaxies are unsuitable environments for this approach as, due to their low metallicity, the CO emission lines are usually not detectable or, if detected, they are faint and with low signal-to-noise ratio \citep[see e.g.][]{1987Tacconi,2009Leroy,2011Bolatto,2012Schruba,2013Elmegreen,2017Cormier,2014Hunt,2015Hunt,2019Hunter}. 
In addition, different methods to estimate $\alpha_\mathrm{CO}$ in dwarf galaxies yield discordant values \citep[e.g.][]{2013Bolatto,2015Hunt,2016Amorin,2019MaddenCormier}, hence it is not clear whether the low luminosity of CO lines represents an intrinsic deficiency of molecular gas or a lack of CO due to the low metallicity. 
Similarly to dwarf galaxies, the outskirts of spiral galaxies are low-metallicity regions where the emission of molecular gas tracers is not detected, but the formation of new stars is actually ongoing \citep[e.g.][]{2007Thilkera,2007Thilkerb,2010Bigiel}. 
Interestingly, some theoretical models show that, in conditions of extremely low metallicity, star formation can directly occur from atomic gas, without pre-existing molecular gas \citep[e.g.][]{2012KrumholzHI,2012GloverClark}. 
Dwarf galaxies are therefore of primary interest in the study of star formation law in order to assess the existence of a unique relation valid for both the low-density and the high-density regions of galaxies. 

While surface-based star formation laws are easily accessible to observations, it is important to ask if they are the manifestation in projection of a more fundamental star formation law based on volume densities such as the one originally proposed by \cite{1959Schmidt}. 
If the latter case is true, it is then possible that some of the variations observed among surface-based star formation laws can be explained as a consequence of varying disc thickness, which depends on the depth of the local gravitational potential \citep[see also][]{2015Elmegreen,2018Elmegreen}. 
\citetalias{2019Bacchini} and \citetalias{2019Bacchini_b} showed that the disc flaring is significant and has an important effect on the observed gas distribution for a sample of ten spiral galaxies, two dwarf galaxies (with moderately-high mass), and the MW. 
In particular, the gas disc was assumed to be in hydrostatic equilibrium in order to derive the radial profile of its scale height. 
This latter was then used to convert the observed radial profiles of the gas and the SFR surface densities to the corresponding volume densities, in order to obtain the volumetric star formation (VSF) law (see \citealt{2008Abramova,2012Barnes,2020Yim} for either similar or different methodologies to derive volumetric relations). 
\citetalias{2019Bacchini} and \citetalias{2019Bacchini_b} found that the VSF law involving the total gas (HI+H$_2$) and the SFR volume densities is a power-law relation with index $\alpha \approx 2$, no break, and a smaller intrinsic scatter than the SK law. 
These authors also discovered a new correlation between the atomic gas and the SFR volume densities, which is steeper (index $\beta \approx 2.8$) than the VSF law but almost equally tight. 

Gas discs in dwarf galaxies, given the shallow gravitational potential, are expected to be thick and significantly flaring in their outskirts 
\citep[see also][]{2010Roychowdhury,2011Banerjee,2018Iorio,2020Patra}. 
Therefore, in order to measure the intrinsic gas distribution of a galaxy, it is mandatory to take into account the radial increase of its scale height. 
There is the compelling possibility that the inefficiency of these galaxies at forming stars may be physically explained as a natural consequence of an ``underlying'' and fundamental VSF law subject to the projection effects due to the flaring. 
Moreover, extending the VSF law to the low-mass regime could allow us to answer key open questions: is the VSF law still valid in the low-density, low-metallicity, and HI-dominated environments? 
Is there a volume density threshold for star formation? 
What is the role of atomic gas in star formation? 

This paper is organised as follows. 
Section~\ref{sec:method} explains the main assumptions of our approach to derive the volume densities based on the hydrostatic equilibrium, describing also the adjustments required for dwarf galaxies. 
Section~\ref{sec:obs} focuses on the sample of dwarf galaxies analysed in this work and the observations available in the literature. Section~\ref{sec:results} presents the main results, which include the atomic gas scale heights and the VSF law for our sample. 
We discuss our findings in Sect.~\ref{sec:discussion} and draw the conclusions in Sect.~\ref{sec:conclusions}. 

\section{Method}\label{sec:method}
In order to derive the volume densities, we adopted the same approach as in \citetalias{2019Bacchini}, which is based on the assumption of hydrostatic equilibrium for the gas disc. 
In this section, we recall the basic assumptions and equations involved; we also discuss the differences in the specific case of dwarf galaxies. 

\subsection{3D distribution of the gas in hydrostatic equilibrium}\label{sec:method_definitions}
We modelled a given galaxy as a system which is symmetric with respect to both its rotation axis (i.e. axisymmetry) and to its midplane (i.e. $z=0$). Hence, its gravitational potential is $\Phi=\Phi(R,z)$, where $R$ is the galactocentric radius. 
We assumed that the gas (either atomic or molecular) is in vertical hydrostatic equilibrium in the galactic potential, meaning that the gas distribution is set by the balance between the gas pressure, which tends to ``inflate'' the gaseous disc, and the gravitational force which pulls the gas towards the midplane. 
The gas pressure is $P=\rho \sigma^2$, where $\rho$ is the gas volume density and $\sigma$ is the gas velocity dispersion. 
For simplicity, we assumed that the velocity dispersion is isotropic and the gas is vertically isothermal, thus $\sigma = \sigma(R)$. 
The vertical gas distribution is then (see e.g. \S~4.6.2 in \citealt*{2019TheBook_CFN})
\begin{equation}\label{eq:rho_Rz_HE}
 \rho(R,z) = \rho(R,0) \exp{\left[ - \frac{\Phi(R,z)-\Phi(R,0)}{\sigma(R)^2} \right]} \, ,
\end{equation}
where $\rho(R,0)$ and $\Phi(R,0)$ are respectively the gas volume density and the total gravitational potential evaluated in the midplane. 
The gas scale height $h$ is defined as the standard deviation of the distribution given by Eq.~\ref{eq:rho_Rz_HE} and its radial profile can be calculated once $\sigma$ and $\Phi$ are known. 

By approximating Eq.~\ref{eq:rho_Rz_HE} near the midplane and integrating along the vertical direction \citep[e.g.][]{1995Olling,2009Koyama}, the radial profile of the volume density in the midplane can be obtained from the surface density $\Sigma(R)$ and the scale height given the relation $\Sigma(R) = \sqrt{2 \pi} \rho(R,0) h(R)$, which is valid for a Gaussian profile with standard deviation $h(R)$. 
This allow us to define the two quantities required to derive the VSF law. 
The first is the volume density of the total gas
\begin{equation}\label{eq:rho_gas}
\begin{split}
 \rho_\mathrm{gas}(R,0)	& = \rho_\mathrm{HI}(R,0) + \rho_\mathrm{H_2}(R,0) \\
			& = \frac{\Sigma_\mathrm{HI}(R)}{\sqrt{2 \pi} h_\mathrm{HI}(R)} + \frac{\Sigma_\mathrm{H_2}(R)}{\sqrt{2 \pi} h_\mathrm{H_2}(R)}\, ,
\end{split}
\end{equation}
where $\Sigma_\mathrm{HI}$ and $h_\mathrm{HI}$ are the surface density and scale height of the atomic gas, and $\Sigma_\mathrm{H_2}$ and $h_\mathrm{H_2}$ are the same but for the molecular gas
\footnote{We use the notation ``HI'' and ``H$_2$'' to indicate the total atomic gas and the total molecular gas, meaning that the surface densities include, when needed, a multiplicative factor of 1.36 to account for the Helium fraction.}. 
In the case of dwarf galaxies, where the atomic gas is likely the dominant gas component, Eq.~\ref{eq:rho_gas} reduces to $\rho_\mathrm{gas} \approx \rho_\mathrm{HI}$. 
The second ingredient for the VSF law is the volume density of the SFR
\begin{equation} \label{eq:rho_SFR}
 \rho_\mathrm{SFR}(R,0) = \frac{\Sigma_\mathrm{SFR}(R)}{\sqrt{2 \pi} h_\mathrm{SFR}(R)} \, ,
\end{equation}
where $\Sigma_\mathrm{SFR}$ and $h_\mathrm{SFR}$ are the SFR surface density and scale height, respectively. 
In \citetalias{2019Bacchini}, two possible definitions of $h_\mathrm{SFR}$ were explored, one constant (100 pc) with the galactocentric radius and the other obtained as the mean of the atomic and molecular gas scale heights weighted by their mass fractions 
\begin{equation}\label{eq:hSFR}
h_\mathrm{SFR}(R) \equiv f_\mathrm{HI}(R) h_\mathrm{HI}(R) + f_\mathrm{H_2}(R) h_\mathrm{H_2}(R) \, , 
\end{equation}
where $f_\mathrm{HI} = \Sigma_\mathrm{HI}/\Sigma_\mathrm{gas}$ and $f_\mathrm{H_2} = \Sigma_\mathrm{H_2}/\Sigma_\mathrm{gas}$. 
Eq.~\ref{eq:hSFR} is based on the idea that the SFR follows the same vertical distribution as the parent gas and gives a radially increasing profile (when the HI and the H$_2$ distributions flare with the galactocentric radius).
\citetalias{2019Bacchini_b} verified that Eq.~\ref{eq:hSFR} is a better choice for the MW with respect to the constant scale height \citep[see also][]{2008Abramova,2012Barnes,2017Sofue}: $h_\mathrm{SFR}(R)$ resulting from Eq.~\ref{eq:hSFR} is in excellent agreement with the scale height of the distributions of classical Cepheids and other tracers of recent star formation in our Galaxy. 
Indications of the flaring of the SFR vertical profile were also found by other authors using mid-infrared observations (i.e. dust-obscured star formation) of nearby galaxies seen at very high inclination angles \citep[e.g.][]{2020Yim,2020ElmegreenElmegreen}. 
We note that, under the assumption that dwarf galaxies are HI-dominated, Eq.~\ref{eq:hSFR} leads to $h_\mathrm{SFR}(R) = h_\mathrm{HI}(R)$. 

\subsection{The mass distribution of dwarf galaxies}\label{sec:method_potential}
The total mass distribution of a star-forming galaxy includes the dark matter (DM) halo, the stellar disc, and the gaseous components (i.e. atomic gas and molecular gas). 
Dwarf galaxies differ from spiral ones in three important aspects. 
First, the DM halo is the dominant mass component at all radii and the shape of the rotation curve indicates that the DM density profile has a core in the centre (e.g. \citealt{2002DeBlok,2011Oh}; but see also e.g. \citealt{2001VandenBosh,2003Swaters} for a different perspective). 
We therefore chose the DM halo model proposed by \cite{2016Read_b,2016Read_c}, which is a Navarro-Frenk-White \citep[NFW; ][]{1996NavarroFrenkWhite} profile with a central core (hereafter core-NFW, see Sect.~\ref{sec:method_potential_DM} for details). 
The second difference between dwarf and spiral galaxies is related to the fact that CO emission is very faint or absent in low-metallicity gas, hence the amount of molecular gas in dwarf galaxies is very uncertain \citep[e.g.][]{2012Schruba,2013Bolatto,2019MaddenCormier}. 
There are indications that the mass fraction of molecular gas is indeed very low compared to the HI mass in dwarf galaxies \citep[$\lesssim 20-40$\%][]{2015Hunt,2019Hunter} and that, even in the extreme case of starbursting dwarfs, the H$_2$ contribution to the potential (traced by the rotation curve) should not exceed that of the stellar disc \citep[e.g.][]{2012Lelli,2014Lelli}. 
Hence, we did not include the molecular gas distribution in the gravitational potential of the dwarf galaxies in this study. 
In Sect.~\ref{sec:discussion_uncertainty_molgas}, we discuss the possible impact on the volume density estimate due to the presence of a fixed fraction of molecular gas. 
The third difference with respect to spiral galaxies is that the contribution of the gas (HI) self-gravity to the gravitational potential of dwarf galaxies is comparable to or higher than that of the stellar disc. 
Therefore, if the gas self-gravity is neglected, the resulting gas scale height is overestimated (see Fig.~A.1 in \citetalias{2019Bacchini}). 
Crucially, as done in \citetalias{2019Bacchini} and \citetalias{2019Bacchini_b}, the gas scale height was calculated numerically using an iterative algorithm which accounts for the gas self-gravity, thus deriving the gas distribution in a self-consistent way (see Sect.~\ref{sec:results_scaleheight}). 

The mass components of our dwarf galaxies are based on \cite{2017Read}, who included the core-NFW DM halo, the stellar disc, and the atomic gas disc. 
In the following, we describe in detail the density profiles of these components, which depend on a set of free parameters. 
In particular, \cite{2017Read} derived these parameters through the decomposition of the HI rotation curves from \cite{2017Iorio}. 
We note that the mass models used in \citetalias{2019Bacchini} were obtained with the same technique.

\subsubsection{The dark matter halo}\label{sec:method_potential_DM}
As mentioned above, the DM density distribution is modelled using the NFW profile modified to account for a central core (i.e. core-NFW) from  \cite{2016Read_b,2016Read_c}. 
This core-NFW profile was obtained using high-resolution numerical simulations of dwarf galaxies with different halo masses evolving in isolation for 14~Gyr. 
Initially, these galaxies have a DM halo with a cusped NFW profile and the cosmological baryon fraction. 
Then, stellar feedback ``heats'' the DM by injecting energy and momentum through supernovae, stellar winds, and radiation pressure, and the cusped profile is gradually transformed into a cored one \citep[see also][]{2005Read,2014Pontzen}
\footnote{For the aim of this work, the exact mechanism which formed the core of the DM halo can be considered irrelevant, as the key point is adopting a functional form for the DM density distribution which can reproduce the observed rotation curve of the dwarf galaxies. Cored DM profiles are also obtained with other mechanisms, such as dynamical friction heating from gas clumps \citep[e.g.][]{2015Nipoti}. }. 
The galaxies resulting from these simulations share similar properties with observed galaxies (i.e. stellar light profile, star formation history, metallicity distribution, gas kinematics) and the density profile of their DM halo is well described by
\begin{equation}\label{eq:rho_cNFW}
 \rho_\mathrm{cNFW}(r) = f^n \rho_\mathrm{NFW} + \frac{n f^{n-1} \left( 1 - f^2 \right)}{ 4 \pi r^2 r_\mathrm{c} } M_\mathrm{NFW} \, ,
\end{equation}
where $r$ is the spherical radius ($r=\sqrt{R^2+z^2}$ in cylindrical coordinates), $\rho_\mathrm{NFW}$ and $M_\mathrm{NFW}$ are respectively the standard NFW halo density and mass profile \citep{1996NavarroFrenkWhite}, $r_\mathrm{c}$ is the core radius, and $f$ is a function than regulates the shape and the extent of the core, defined as
\begin{equation}\label{eq:fn}
 f = \tanh \left( \frac{r}{r_\mathrm{c}} \right)  \, ,
\end{equation}
while $0 < n \leq 1$ ($n=0$ corresponds to no core). 
The parameter $n$ is given by
\begin{equation}
 n=\tanh \left( \kappa \frac{t_\mathrm{SF}}{t_\mathrm{dyn}} \right) \, ,
\end{equation}
where
\begin{equation}
 t_\mathrm{dyn} = 2 \pi \sqrt{ \frac{ r_\mathrm{s}^3 }{ G M_\mathrm{NFW}(r_\mathrm{s}) }}
\end{equation}
is the orbital time at the scale radius $r_\mathrm{s}$ of the NFW profile, $t_\mathrm{SF}=14$~Gyr is the total star formation time, and $\kappa = 0.04$. 
\cite{2017Read} found $r_\mathrm{c}=2.94 R_\star$ with $R_\star$ being the stellar disc scale length. 

\subsubsection{Disc components}\label{sec:method_potential_disc}
The stellar disc mass distribution is modelled with an exponential radial profile and a $\sech^2$ vertical profile \citep{1981VanderKruit_a},
\begin{equation}\label{eq:expsec_stardisc}
 \rho_\star (R,z) = \rho_{\star,0} \exp \left( - \frac{R}{R_\star} \right) \sech^2 \left(\frac{z}{z_\star} \right) \, ,
\end{equation}
where $\rho_{\star,0}=\Sigma_{\star,0}/( 2 z_\star)$ is the central density, $R_\star$ is the stellar scale length, and $z_\star$ the scale height, 
which is assumed to be equal to $R_\star/5$ (see \citealt{2011vanderKruitFreeman} and references therein). 

We modelled the atomic gas disc using a combination of a fourth-order polynomial and an exponential function
\begin{equation}\label{eq:poly}
  \Sigma_\mathrm{HI} (R) = \Sigma_\mathrm{HI,0} \left( 1 + C_1 R + C_2 R^2 + C_3 R^3 + C_4 R^4 \right) 
			   \exp \left( -\frac{R}{R_\Sigma} \right) \,,
\end{equation}
where $\Sigma_\mathrm{HI,0}$ is the central surface density, $R_\Sigma$ is the scale radius, and $C_i$ are the polynomial coefficients. 
We choose Eq.~\ref{eq:poly} as it is flexible and can easily model the variety of observed profiles of $\Sigma_\mathrm{HI}(R)$ (see Sect.~\ref{sec:obs_HI}). 
The HI distribution modelled using this functional form also serves to improve the efficiency of the numerical computation of HI gravitational potential and scale height (see Sect.~\ref{sec:results_scaleheight}). 

\subsection{The gas velocity dispersion}\label{sec:method_vdisp}
The radial profile of the velocity dispersion $\sigma_\mathrm{HI}(R)$ is modelled by an exponential function 
\begin{equation}\label{eq:exp_vdisp}
 \sigma_\mathrm{HI}(R) = \sigma_\mathrm{HI,0} \exp \left( -\frac{R}{R_{\sigma}} \right) \, ,
\end{equation}
where $\sigma_\mathrm{HI,0}$  is the velocity dispersion at the galaxy centre and $R_{\sigma}$ is a scale radius. 
This functional form can satisfactorily model the shape of the radial profile of the HI velocity dispersion typically observed in nearby galaxies, including a linear decline or a constant value (\citealt{2008Boomsma,2009Tamburro,2017Marasco,2017Iorio}; \citetalias{2019Bacchini}). 
As in the case of $\Sigma_\mathrm{HI}$, the use of an analytic $\sigma_\mathrm{HI}(R)$ is purely for the sake of improving the computational speed of the numerical integration to derive $\Phi$ and $h_\mathrm{HI}$.

\section{Sample description}\label{sec:obs}
We selected a sample of dwarf galaxies with the following information available in the literature: i) robust HI kinematics (i.e. velocity dispersion and rotation curve); ii) parametric models of the mass distribution (see Sect.~\ref{sec:method_potential}); iii) azimuthally averaged radial profiles of HI and SFR surface densities. 
A sub-sample of ten galaxies from the Local Irregulars That Trace Luminosity Extremes, The HI Nearby Galaxy Survey \citep[LITTLE THINGS;][]{2012Hunter} fulfills these requirements, as we explain in the following. 

\subsection{Atomic gas kinematics and surface density}\label{sec:obs_HI}
\cite{2017Iorio} analysed the HI kinematics for 17 LITTLE THINGS galaxies using the software $^{\mathrm{3D}}$\textsc{Barolo}
\footnote{\url{http://editeodoro.github.io/Bbarolo/}} \citep{2015Diteodoro}. 
This software performs a tilted-ring model fitting using the emission-line data cube of a galaxy and derives the geometric parameters (e.g. inclination, position angle, kinematic centre) and the kinematics (e.g. rotation curve, velocity dispersion, systemic velocity) of the gas disc. 
In addition, \cite{2017Iorio} included a robust estimate for the asymmetric drift correction to the gas circular speed. 
This correction is fundamental in the case of dwarf galaxies, where pressure support can be significant. 
In our work, we excluded the galaxies with disturbed HI kinematics (i.e. NGC~1569) and those with either asymmetric drift-dominated or very uncertain rotation curves (i.e. CVnIdwA, DDO~53, DDO~210, DDO~216,UGC~8508). 
In addition, DDO~154 was removed from the sample as it was studied already in \citetalias{2019Bacchini}. 
We thus obtained a sample of ten dwarf galaxies, whose main properties are reported in Table~\ref{tab:sample}. 
\renewcommand{\arraystretch}{1.5}
\begin{table*}
	\centering
	\caption{Properties of the sample galaxies (see \citealt{2017Iorio} for details): 
		(1) distance; 
		(2) integrated SFR from FUV (see Sect.~\ref{sec:obs_SFRD});
		(3) mean circular velocity (i.e. asymmetric-drift corrected) of the outer HI disc;
		(4) HI disc inclination; 
		(5) HI disc position angle. 
		Parameters of the best-fit model for the core-NFW halo and the stellar disc from \cite{2017Read}: 
		(6) DM mass within $R_\mathrm{200}$; 
		(7) DM halo concentration;
		(8) index of the DM halo core function;
		(9) stellar mass; 
		(10) stellar disc scale lenght.}
	\label{tab:sample}
	\begin{tabular}{l c c c c c c c c c c c}
		\hline\hline
		Galaxy   & D     & log(SFR)                     & $\langle V_\mathrm{c} \rangle$ & $\langle i\rangle$ & $\langle\mathrm{P.A.}\rangle$ & $M_\mathrm{200}$       & $c$                  & $n$   & $M_\star$            & $R_\star$ &  \\
		& (Mpc) & ($\mathrm{M_\odot yr^{-1}}$) & (\kms)                           & ($^{\circ}$)       & ($^{\circ}$)                  & ($10^{10}$ M$_\odot$)  &                      &       & ($10^{7}$ M$_\odot$) & (kpc)     &  \\
		& (1)   & (2)                          & (3)                              & (4)                & (5)                           & (6)                    & (7)                  & (8)   & (9)                  & (10)      &  \\ \hline
		WLM      & 1.0   & $-2.13 \pm 0.08$             & $38.7 \pm 3.4$                   & $74.0 \pm 2.3$     & $174.0 \pm 3.1$               & $0.8^{+0.2}_{-0.2}$   & $17$   & 0.887 & $1.62$               & $0.75$    &  \\
		DDO~52   & 10.3  & $-1.82 \pm 0.08$             & $51.1 \pm 6.3$                   & $55.1 \pm 2.9$     & $6.5 \pm 2.0$                 & $1.20^{+0.29}_{-0.27}$  & $17.3$ & 0.894 & $5.27$               & $0.94$    &  \\
		DDO~87   & 7.4   & $-1.95 \pm 0.07$             & $50.3 \pm 9.1$                   & $42.7 \pm 7.3$     & $238.6 \pm 4.7$               & $1.13^{+0.27}_{-0.25}$ & $17.6$ & 0.900 & $3.3$                & $1.13$    &  \\
		NGC~2366 & 3.4   & $-0.97 \pm 0.07$             & $57.7 \pm 5.4$                   & $65.1 \pm 4.2$     & $39.8 \pm 2.8$                & $2.40^{+0.59}_{-0.54}$  & $17.3$  & 0.894 & $6.95$               & $1.54$    &  \\
		DDO~126  & 4.9   & $-1.80 \pm 0.06$             & $38.6 \pm 3.1$                   & $62.2 \pm 2.9$     & $140.7 \pm 3.5$               & $0.6^{+0.2}_{-0.1}$   & $20.6$ & 0.947 & $1.61$               & $0.82$    &  \\
		DDO~168  & 4.3   & $-1.67 \pm 0.06$             & $56.2 \pm 6.9$                   & $62.0 \pm 2.6$     & $272.7 \pm 4.2$               & $2.10^{+0.52}_{-0.48}$  & $16.8$ & 0.883 & $5.9$                & $0.82$    &  \\
		DDO~133  & 3.5   & $-1.91 \pm 0.05$             & $47.2 \pm 5.1$                   & $38.9 \pm 3.7$     & $-3.8 \pm 8.9$                & $1.60^{+1.10}_{-0.44}$   & $25.8$ & 0.984 & $3.04$               & $0.804$   &  \\
		DDO~50   & 3.4   & $-0.95 \pm 0.04$             & $38.7 \pm 10.1$                  & $33.1 \pm 4.6$     & $174.9 \pm 5.9$               & $0.32^{+0.16}_{-0.08}$ & $26.2$ & 0.985 & $10.72$              & $0.89$    &  \\
		DDO~47   & 5.2   & $-1.61 \pm 0.04$             & $62.6 \pm 5.2$                   & $37.4 \pm 1.7$     & $317.4 \pm 7.3$               & $4.4^{+3.2}_{-1.9}$    & $20.5$ & 0.946 & $9.4$                & $0.7$     &  \\
		DDO~101  & 6.4   & $-2.45 \pm 0.09$             & $52.9 \pm 3.6$                   & $52.4 \pm 1.7$     & $287.4 \pm 3.6$               & $5.2^{+0.6}_{-0.4}$    & $28.9$ & 0.993 & $6.54$               & $0.58$    &  \\ \hline
	\end{tabular}
\end{table*}

\begin{figure*}
	\includegraphics[width=2.\columnwidth]{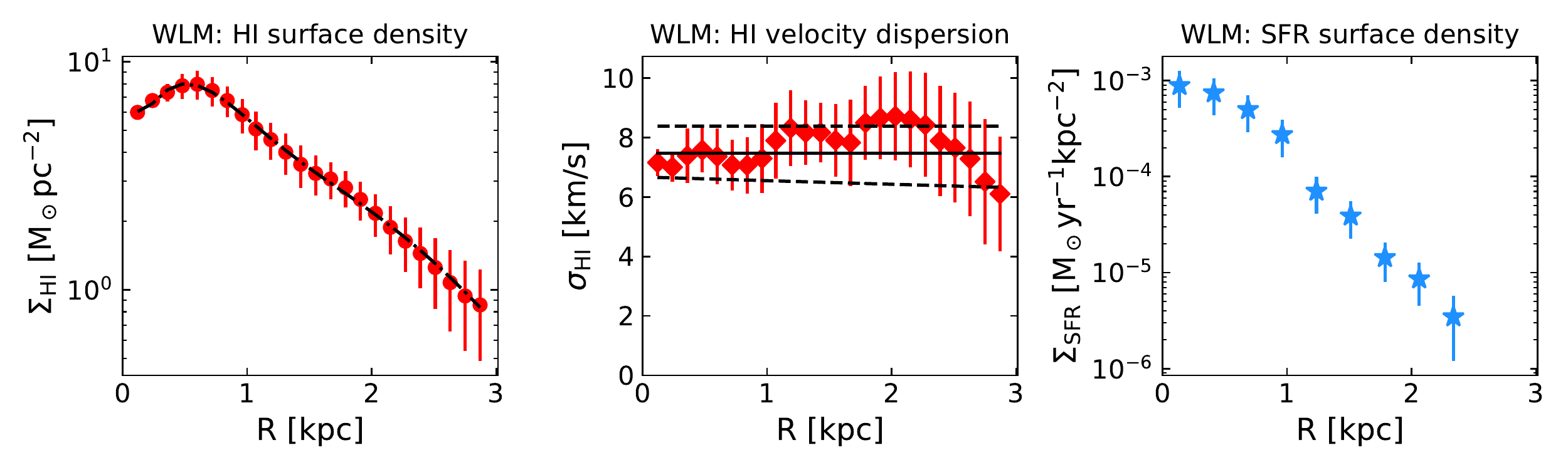}
	\caption{Azimuthally averaged radial profiles of the atomic gas surface density (left panel) and the HI velocity dispersion (central panel) from \cite{2017Iorio}, and of the SFR surface density (right panel) from FUV photometry \citep{2012Zhang} for WLM. 
		The dot-dashed black curve in the left panel is the best-fit polynomial function (Eq.~\ref{eq:poly}) used to model the gas radial distribution. 
		The solid and the dashed black lines in the central panel are the best-fit functions (Eq.~\ref{eq:exp_vdisp}) used to model, respectively, the velocity dispersion and its upper and lower limits, which were adopted to derive the scale height and its uncertainty. }
	\label{fig:wlm_sdhi_vdhi_sfrd}
\end{figure*}

As already mentioned, the rotation curves of the dwarf galaxies in \cite{2017Iorio} were used by \cite{2017Read} to obtain the mass models, which include the core-NFW DM halo described in Sect.~\ref{sec:method_potential_DM}, a stellar disc, and an atomic gas disc. 
Both disc components were assumed razor-thin and with an exponentially declining radial profile, hence we substituted these models with those described in Sect.~\ref{sec:method_potential_disc}. 
We used the same stellar mass and scale length as \cite{2017Read}, which are reported in Table~\ref{tab:sample} together with the parameters for the DM halo model. 
For the atomic gas disc, we derived $\Sigma_\mathrm{HI,0}$, $C_i$, and $R_\mathrm{HI}$ by fitting Eq.~\ref{eq:poly} to the observed surface density of HI from \cite{2017Iorio}. 
The left panel in Fig.~\ref{fig:wlm_sdhi_vdhi_sfrd} shows the azimuthally averaged radial profile of the atomic gas surface density for WLM from \cite{2017Iorio}, and the polynomial best-fit model (dot-dashed black curve). 
In Appendix~\ref{ap:obs_ALL}, we show $\Sigma_\mathrm{HI}(R)$ and the polynomial fits for the rest of the sample. 
We note that the profiles typically differ from an exponential and have diverse shapes. 
We verified that the updated atomic gas radial profile and stellar scale height do not imply significant variations in the resulting circular velocity generated by the total mass distribution of our galaxies, as a consequence of these systems being largely dark-matter dominated. 

We used the radial profiles of the HI velocity dispersion derived by \cite{2017Iorio} to fit Eq.~\ref{eq:exp_vdisp} and obtain the model for $\sigma_\mathrm{HI}(R)$. 
We recall that this step aims to improve the computational speed in the iterative calculation of the scale height, but it does not influence the resulting $h_\mathrm{HI}(R)$. 
The central panel in Fig.~\ref{fig:wlm_sdhi_vdhi_sfrd} shows, for WLM, the observed velocity dispersion of HI (red diamonds) and the best-fit exponential profile (black solid curve). 
The dashed black curves are the fits to $\sigma_\mathrm{HI} \pm \Delta \sigma_\mathrm{HI}$, where $\Delta \sigma_\mathrm{HI}$ is the uncertainty, and were used to obtain the uncertainties on the atomic gas scale height (see Sect.~\ref{sec:results_scaleheight}). Figure~\ref{fig:all_sdhi_vdhi_sfrd_1} shows the profiles for the other galaxies in the sample. 
We note that the velocity dispersion is approximately constant for some galaxies (e.g. WLM, DDO~47), but in general $\sigma_\mathrm{HI}(R)$ declines with increasing galactocentric radius. 

\subsection{Star formation rate surface density}\label{sec:obs_SFRD}
In order to derive the SFR surface density for our galaxies, we used the azimuthally averaged surface photometry from \cite{2012Zhang}, based on far-ultraviolet (FUV) and near-ultraviolet (NUV) images obtained with the Galaxy Evolution Explorer \citep[GALEX,][]{2005Martin}. 
We calculated the FUV magnitude per square arcsecond as $\mu_\mathrm{FUV} = \mu_\mathrm{NUV} + (\mathrm{FUV-NUV})$, where $\mu_\mathrm{NUV}$ is the NUV AB absolute magnitude per square arcsecond and $(\mathrm{FUV-NUV})$ is the FUV-NUV color corrected assuming \cite{1989Cardelli} dust extinction law (see \citealt{2012Zhang} for details)  
\footnote{We note that the dust extinction correction is typically minimal for our dwarf galaxies \citep[see][]{2010Hunter,2012Zhang}.}. 
The corresponding FUV surface brightness is
\begin{equation}\label{eq:F_FUV}
 F_\nu(\mathrm{FUV}) = 10^{- 0.4 \left[ \left( \frac{\mu_\mathrm{FUV}}{\mathrm{mag \, arcsec^{-2}}} \right) +48.6 \right]} \, \mathrm{erg~s^{-1} cm^{-2} Hz^{-1} arcsec^{-2}} \, .
\end{equation}
In order to estimate the SFR surface density from the FUV surface brightness, we adopted the calibration of the SFR-FUV relation by \cite{2015McQuinn}. 
This calibration is ideally suitable for our study, as it was obtained for a sample of nearby dwarf galaxies taking into account their low metallicity and a stochastically populated initial mass function (IMF). 
These authors derived the SFR from resolved stellar populations by fitting the color-magnitude diagrams (CMDs) with synthetic stellar populations generated from stellar evolution libraries and used this SFR to calibrate the SFR-FUV relation. 
We adapted \cite{2015McQuinn}'s relation from a Salpeter IMF to a Kroupa IMF by including a multiplicative factor of 0.67 \citep{2012KennicuttEvans,2014MadauDickinson}. 
This correction was applied for consistency with \citetalias{2019Bacchini}, where the SFR surface densities were based on a Kroupa IMF \citep[see][]{2008Leroy}. 
This gives the following relation between the FUV luminosity $L_\nu(\mathrm{FUV})$ and the SFR
\begin{equation}\label{eq:SFR_FUV}
 \mathrm{SFR} = (1.37 \pm 0.54) \times 10^{-28} \left[ \frac{ L_\nu(\mathrm{FUV})}{\mathrm{erg~s^{-1} Hz^{-1}}} \right] \, \mathrm{M_\odot yr^{-1}} \, .
\end{equation}
As pointed out by \cite{2015McQuinn}, Eq.~\ref{eq:SFR_FUV} yields, for the same FUV luminosity, values of SFR that are about $50-70$\% higher that other calibrations in the literature derived for galaxies with higher mass and metallicity than dwarf galaxies \citep[e.g.][]{2011Hao,2007Salim}. 
The SFR surface density can be obtained by combining Eq.~\ref{eq:F_FUV} and Eq.~\ref{eq:SFR_FUV} 
\begin{equation}\label{eq:Sigma_SFR}
 \Sigma_\mathrm{SFR} = \cos i \times 10^{- 0.4 \left( \frac{\mu_\mathrm{FUV}}{\mathrm{mag \, arcsec^{-2}}} \right) + 7.415} \, \mathrm{M_\odot yr^{-1} kpc^{-2}} \, ,
\end{equation}
where the $\cos i$ term corrects for the galaxy inclination (see Table~\ref{tab:sample})
\footnote{Despite the different calibration assumed in this work, the SFR surface density given by Eq.~\ref{eq:Sigma_SFR} approximately corresponds the one reported in \cite{2015ElmegreenHunter} and obtained with the SFR-FUV relation adopted in \cite{2010Hunter}, which assumes a Salpter IMF.}. 
The uncertainty on $\Sigma_\mathrm{SFR}$ was calculated through error propagation in order to take into account the uncertainties on $\mu_\mathrm{NUV}$, the FUV-NUV color, $i$ and the calibration constant in Eq.~\ref{eq:SFR_FUV}. 
The right panel of Fig.~\ref{fig:wlm_sdhi_vdhi_sfrd} shows the radial profile of the SFR surface density for WLM resulting from Eq.~\ref{eq:SFR_FUV}, while the profiles for the rest of the galaxies in our sample are shown in Fig.~\ref{fig:all_sdhi_vdhi_sfrd_1}. 
In Table~\ref{tab:sample}, we report the SFR integrated over the whole star-forming disc of our galaxies. 

We note that the SFR surface densities adopted in \citetalias{2019Bacchini} were taken from \cite{2008Leroy}, who used a combination of FUV and $24 \, \mu$m luminosities in order to take into account the emission from massive and young stars absorbed by dust and re-radiated at the mid-infrared  wavelenghts. 
To convert these luminosities to $\Sigma_\mathrm{SFR}$, these authors assumed a relation obtained for normal star-forming galaxies, which may not be applicable to dwarf galaxies \citep{2011Hao}. 
However, in Sect.~\ref{sec:discussion_uncertainty_obscuredSF}, we explore the effect of using \cite{2008Leroy} prescription to derive the SFR surface density for our sample of galaxies. 
We anticipate that our results are not significantly affected if we use this correction. 

\section{Results}\label{sec:results}
In this section, we present our findings by focusing first on the scale height of the atomic gas discs of our sample of galaxies and then on the VSF law. 

\subsection{The scale height}\label{sec:results_scaleheight}
We derived the HI scale height using the Python module \textsc{Galpynamics}\footnote{\url{https://github.com/iogiul/galpynamics}} \citep[see][]{2018Iorio}. 
This code calculates the gravitational potential generated by one or more mass components described by a parametric model through numerical integration, allowing to derive useful quantities, such as the circular velocity and the scale height of a gaseous mass component. 
\textsc{Galpynamics} is based on the assumption of vertical hydrostatic equilibrium and includes an iterative algorithm which accounts for the gas self-gravity. 

In the following, we briefly explain how this code operates in the case of our dwarf galaxies. 
The mass components and their parameters are described in Sect.~\ref{sec:method_potential} and Table~\ref{tab:sample}, respectively. 
We recall that the HI velocity dispersion and surface density are modelled using the observations described in Sect.~\ref{sec:obs_HI}. 
As a preliminary step, \textsc{Galpynamics} calculates the external potential $\Phi_\mathrm{ext}$, which is fixed at all the iterations, given by the DM halo (Eq.~\ref{eq:rho_cNFW}) and the stellar disc (Eq.~\ref{eq:expsec_stardisc}) mass distributions. 
The HI disc is initially assumed to be razor-thin and its contribution ($\Phi_\mathrm{HI}$) to the total gravitational potential ($\Phi = \Phi_\mathrm{ext}+\Phi_\mathrm{HI}$) is estimated from the surface density distribution given by Eq.~\ref{eq:poly}. 
Using this preliminary total potential, the gas distribution is calculated through Eq.~\ref{eq:rho_Rz_HE}, where the velocity dispersion is given by Eq.~\ref{eq:exp_vdisp}. 
A first guess of the gas scale height is obtained by fitting a Gaussian profile to gas vertical distribution at each radius and extracting its standard deviation. 
Then, $\Phi$ is updated using the potential generated by a vertically extended gas distribution as found in the previous step, and a new estimate of the scale height is obtained through the Gaussian fitting. 
This procedure is iterated until either the relative residual between two successive determinations of the scale height differ by less than $10^{-4}$ or more than ten iterations are completed. 

\begin{figure*}
\includegraphics[width=2.\columnwidth]{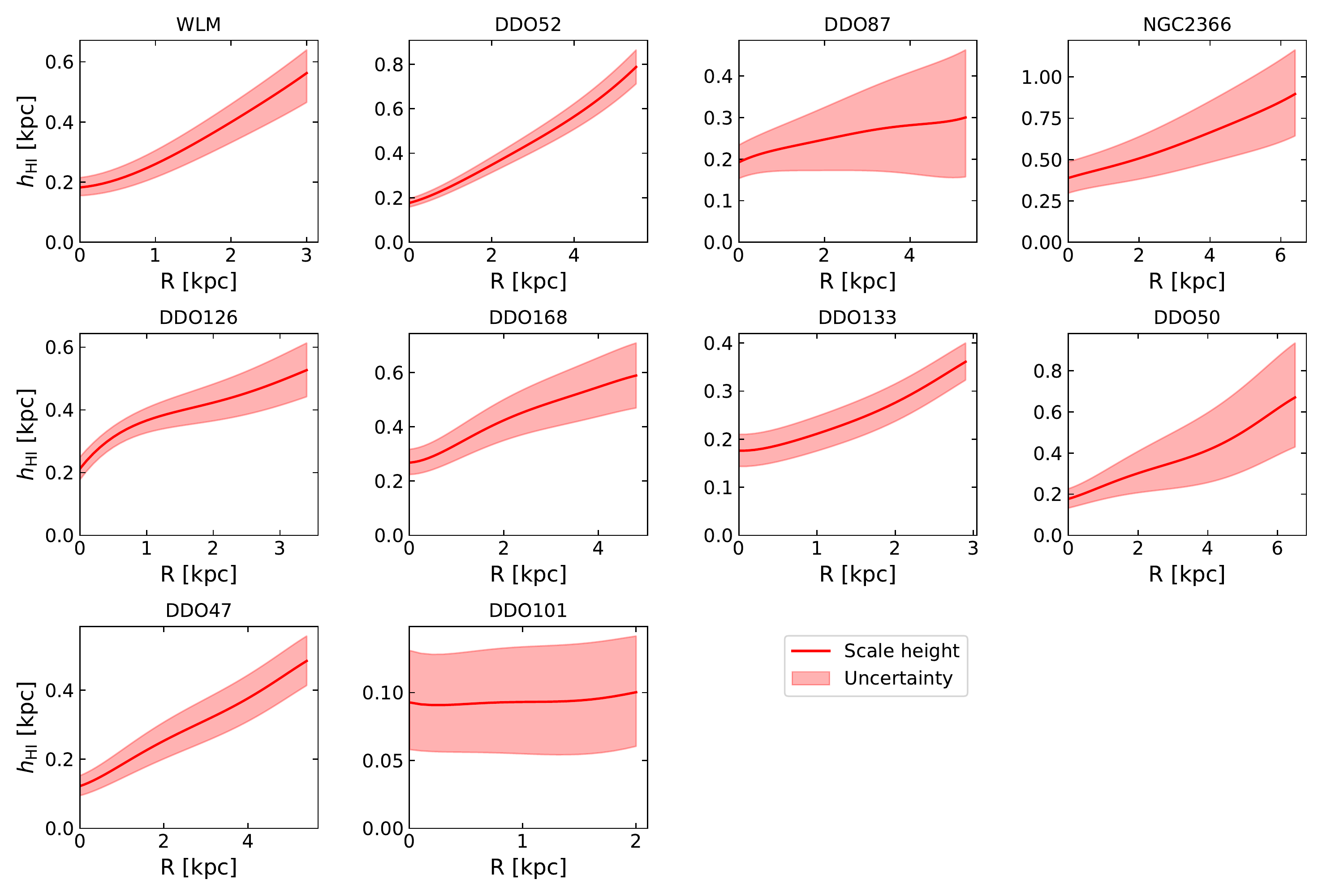}
\caption{Radial profiles of the atomic gas scale height derived under the assumption of vertical hydrostatic equilibrium (red curve). 
The red area shows the uncertainty on $h_\mathrm{HI}(R)$ (see text).}
\label{fig:hHI_all}
\end{figure*}

Figure~\ref{fig:hHI_all} shows the scale height of the atomic gas distribution as a function of the galactocentric radius for the galaxies in our sample. 
Despite the relatively small size of the discs (3-4~kpc on average), the scale heights significantly increase with the distance from the galactic centre, reaching a few hundreds of parsec at the outermost radii. 
The gas distribution flares of a factor ranging from a minimum of $\approx 1.1-1.6$ (but with large uncertainty) for DDO~101 and DDO~87 to a maximum of $\approx 4-4.5$ (e.g. DDO~52, DDO~47). 
The red band represents the uncertainty on the scale height and the lower and the upper limits on $h_\mathrm{HI}$ were calculated taking into account the uncertainties on the HI velocity dispersion (see Sect.~\ref{sec:obs_HI}), as done in \citetalias{2019Bacchini}. 
We have verified that the uncertainties on the DM halo masses and concentrations reported by \cite{2017Read} correspond to at most $\approx 5$\% uncertainty on our scale height estimates, confirming the gas velocity dispersion as the main source of uncertainty in our modelling.

\subsection{The volumetric star formation law}\label{sec:results_VSF}
In order to assess the validity of the volumetric relations obtained in \citetalias{2019Bacchini}, we derived the volume densities for our sample of dwarf galaxies through Eq.~\ref{eq:rho_gas} and Eq.~\ref{eq:rho_SFR}, using the HI scale heights calculated in Sect.~\ref{sec:results_scaleheight} and the surface density profiles described in Sect.~\ref{sec:obs_HI} and Sect.~\ref{sec:obs_SFRD}. 
We note that the conversion to volume densities is not simply a shift in the normalisation of the radial profile of the surface density, as each point is divided by a different scale height. 
In other words, the radial gradient of the volume density profile is different from that of surface density profile because of the disc flaring.

We recall that the sample of spiral galaxies studied in \citetalias{2019Bacchini} has a significant fraction of molecular gas, which is typically dominant in the innermost regions (i.e. $f_\mathrm{H_2} \approx 1$) but essentially negligible at larger radii. 
In the outskirts, the total gas volume density is dominated by the atomic gas volume density and $f_\mathrm{HI} \gg f_\mathrm{H_2}$. 
Accordingly, Eq.~\ref{eq:hSFR} results in $h_\mathrm{SFR}(R)$ which is very similar or close to $h_\mathrm{H_2}$ near the galactic centre and gradually turns into $h_\mathrm{HI}$ with increasing radius. 
In \citetalias{2019Bacchini_b}, the scale height of the distribution of classical Cepheids was used to calculate the SFR volume density in the MW. 
In addition, it was verified that the scale height of recent star formation tracers (e.g. classical Cepheids, OB stars) is very similar to that of the gas in hydrostatic equilibrium calculated using Eq.~\ref{eq:hSFR} out to $R \approx 20$~kpc. 
In the case of dwarf galaxies, the atomic gas is likely the dominant gaseous component, therefore we chose to assume $f_\mathrm{H_2} \approx 0$ as a fiducial case of study, which implies that Eq.~\ref{eq:rho_gas} and Eq.~\ref{eq:hSFR} respectively reduce to $\rho_\mathrm{gas} = \rho_\mathrm{HI}$ and $h_\mathrm{SFR} = h_\mathrm{HI}$.
We explore the effect of including a fraction of molecular gas in Sect.~\ref{sec:discussion_uncertainty_molgas}. 

Dwarf galaxies share similarities with the outskirts of spiral galaxies, as both these environments are metal poor and HI-dominated. 
Hence, we extended the analysis on the galaxy sample of \citetalias{2019Bacchini} by including also the regions beyond the stellar disc (i.e. $R >R_{25}$) where the SFR surface density is observed. 
We adopted the $\Sigma_\mathrm{SFR}$ from \cite{2010Bigiel}, who used the same GALEX FUV maps and SFR-FUV relation as \cite{2008Leroy} but measured of the SFR surface density up to larger radii. 
The atomic gas surface density and scale height of the outermost star-forming regions were taken from \cite{2020Bacchini}. 
In particular, $\Sigma_\mathrm{HI}$ was obtained from 21-cm data cubes using $^{\mathrm{3D}}$\textsc{Barolo} and $h_\mathrm{HI}$ was derived with the same method based on the hydrostatic equilibrium as done in \citetalias{2019Bacchini} (see also Sect.~\ref{sec:results_scaleheight}). 
This allowed us to study the VSF law up to larger radii with respect to \citetalias{2019Bacchini}: we could extend the profiles by $\sim 1$~kpc for 5 galaxies (DDO~154, IC~2574, NGC~0925, NGC~4736, and NGC~7793), while for NGC~2403 and NGC~3198 we increased the radial coverage by $\approx 5$~kpc and $\approx 10$~kpc, respectively. 

In the following, we verify whether the correlations obtained in \citetalias{2019Bacchini}, namely the VSF law between the total gas and the SFR volume densities and the relation including the atomic gas only, are valid for the new sample of dwarf galaxies and the outermost star-forming regions of the galaxies in \citetalias{2019Bacchini}. 

\subsubsection{The VSF law with total gas}\label{sec:results_VSF_gas}
\begin{figure*}
	\includegraphics[width=2.\columnwidth]{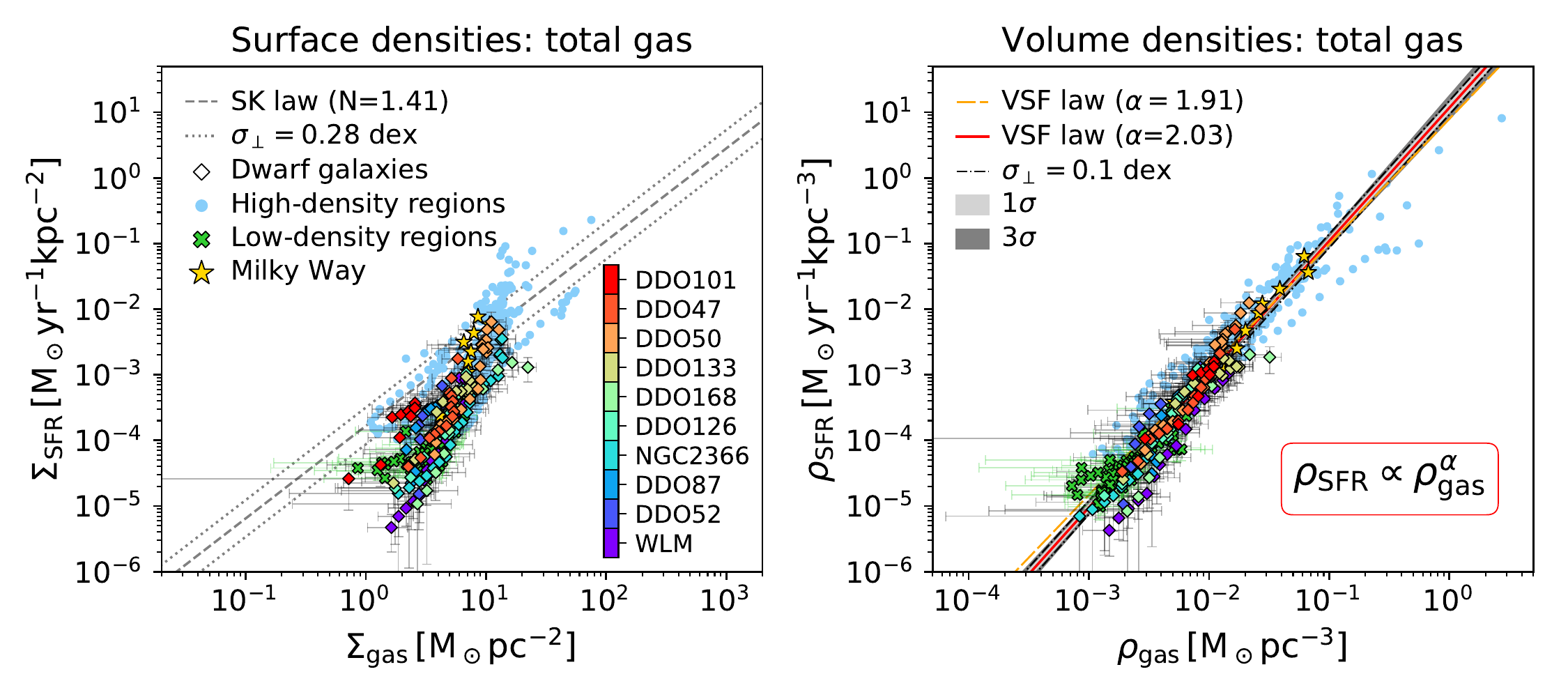}
	\caption{Star formation laws based on the surface densities (left) and the midplane volume densities (right) of the total gas and the SFR for the sample of ten dwarf galaxies studied in this paper (diamonds). 
		Each color indicates the azimuthally averaged radial profile of a different galaxy according to the color-bar. 
		The light blue points and the green crosses respectively indicate the star-forming regions within and beyond the optical radius for the sample of 12 galaxies of \citetalias{2019Bacchini}, while the yellow stars are for the MW (\citetalias{2019Bacchini_b}). 
		In the left panel, the grey dashed line is the SK law from \cite{2019delosReyesKennicutt} with its intrinsic scatter (grey dotted lines). 
		In the right panel, the long-dashed orange line shows the best-fit VSF law from \citetalias{2019Bacchini}, while the red solid line is the best-fit obtained in this work. 
		The black dot-dashed line displays the orthogonal intrinsic scatter of the new best-fit, while the grey areas indicates the $1\sigma$ and $3\sigma$ uncertainties on the fit. 
		The VSF law has a smaller scatter than the SK law and it is a power-law with index $\alpha \approx 2$, showing no indication for a break over a wide range of densities.}
	\label{fig:vsf}
\end{figure*}

The gas and the SFR surface densities of our sample of dwarf galaxies are shown in the left panel of Fig.~\ref{fig:vsf} (colored diamonds). 
The light blue circles represent the 12 galaxies examined in \citetalias{2019Bacchini} and the yellow stars correspond to the MW (see \citetalias{2019Bacchini_b} for details). 
The colored diamonds indicate the new sample of dwarf galaxies and the green ``x'' correspond to the outermost star-forming regions of seven galaxies from \citetalias{2019Bacchini}'s sample, as explained in the previous section. 
We can see that both dwarf and spiral galaxies detach from the SK law for $\Sigma_\mathrm{gas} \lesssim 10~\mathrm{M_\odot pc^{-2}}$. 
This behaviour in the low-density regime is widely reported in the literature and increases the scatter in the correlation \citep[e.g.][]{2007Kennicutt,2008Bigiel,2011Bolatto,2014Dessauges,2019delosReyesKennicutt}. 
We note that the relation shown in Fig.~\ref{fig:vsf} is the ``global'' SK law \citep{2019delosReyesKennicutt}, hence a small shift (mostly in normalisation) with respect to the spatially resolved surface densities is expected \citep[e.g.][]{2007Kennicutt}. 
However, this shift does not seem sufficient to describe both the regimes of dwarf and spiral galaxies, which appear to loosely follow a broken power-law with $N \approx 1.4$ in the high-density regions and $N \approx 3$ at low densities \citep[see also][]{2013Gatto}. 
We fitted the SK law with $N=1.41$ to the surface densities shown in Fig.~\ref{fig:vsf} using the Python module \texttt{emcee} \citep{2013ForemanMackey}, finding that the normalisation of the best-fit relation is lower ($-4.27 \pm 0.02$) than the value obtained by \citealt{2019delosReyesKennicutt} ($-3.84_{-0.09}^{+0.08}$), while the scatter is similar \citep[$0.30 \pm 0.01$~dex; see also][]{2018Shi}. 
We repeated the fit leaving the index as a free parameter and found that the best-fit relation has index $3.17^{+0.11}_{-0.10}$, normalisation $-5.73 \pm 0.03$ and smaller intrinsic scatter ($0.16 \pm 0.01$~dex) with respect to our best-fit SK law with $N=1.41$. 
We anticipate that these correlations based on the surface densities have a larger scatter than the VSF law.

The right panel of Fig.~\ref{fig:vsf} shows the volume densities for the various samples using the same symbols as in the left panel. 
Remarkably, the dwarf galaxies either closely follow or are compatible within the errors with the VSF law found in \citetalias{2019Bacchini}, extending the validity of the relation to the low-density regime down to $\rho_\mathrm{gas} \sim 10^{-3}~\mathrm{M_\odot pc^{-3}}$. 
Also the range of SFR volume densities is extended downward by one order of magnitude. 
\citetalias{2019Bacchini} obtained that the VSF law between the SFR and the total gas volume densities is described by
\begin{equation}\label{eq:log_vsf_gas}
 \log \left( \frac{\rho_\mathrm{SFR}}{\mathrm{M}_\odot \mathrm{yr}^{-1} \mathrm{kpc}^{-3}} \right) = 
 \alpha
 \log \left( \frac{\rho_\mathrm{gas}}{\mathrm{M}_\odot \mathrm{pc}^{-3}} \right) 
 + \log A \, ,
\end{equation}
with best-fit slope $\alpha=1.91 \pm 0.03$, normalisation $\log A=0.90 \pm 0.02$, and perpendicular scatter $\sigma_\perp = 0.12 \pm 0.01$~dex. 
We repeated the fit including also the sample of dwarf galaxies and the outermost star-forming regions of the sample of \citetalias{2019Bacchini}, finding $\alpha=2.03 \pm 0.03$, $\log A=1.10 \pm 0.01$, and $\sigma_\perp = 0.10 \pm 0.01$~dex. 
This confirms that the relation between the total gas and the SFR volume densities is $\rho_\mathrm{SFR} \propto \rho_\mathrm{gas}^\alpha$ with $\alpha \approx 2$ and that the VSF law has a smaller scatter the SK law. 
We note that the best-fit VSF law obtained in this work and that derived in \citetalias{2019Bacchini} are compatible within $\lesssim 2\sigma$ (considering that the two relations have very similar $1\sigma$ error). 

From a thorough inspection of Fig.~\ref{fig:vsf}, we can see that a few points belonging to three galaxies, DDO~87, DDO~168 and WLM, seem slightly detached from the VSF law. 
We note that DDO~168 and DDO~87 are actually compatible within the errors with the best-fit relation. 
The data points of WLM, instead, appear to be slightly more than $3\sigma$ from the relation, also in light of their relatively small error-bar. 
It is unlikely that this difference is entirely due to inaccuracies in the determination of $\rho_\mathrm{gas}$ as the scale height, the surface density, and their uncertainties are quite robust. 
The profile of the SFR surface density also seems reliable, as it is similar to other estimates in the literature obtained from FUV emission \citep[e.g.][]{2018Mondal}. 
Our $\Sigma_\mathrm{SFR}$ does not take into account the possible contribution of dust-obscured star formation, which is instead included in \cite{2008Leroy} profiles. 
The effect of including the dust-obscured star formation is discussed in detail in Sect.~\ref{sec:discussion_uncertainty_obscuredSF}, but we anticipate that the global SFR of WLM is increased of about 15\% by using the same method as \cite{2008Leroy}, reducing of a factor of $\approx 2$ the gap with the VSF law. 

\subsubsection{Relation with the atomic gas only}\label{sec:results_VSF_HI}
\begin{figure*}
\includegraphics[width=2.\columnwidth]{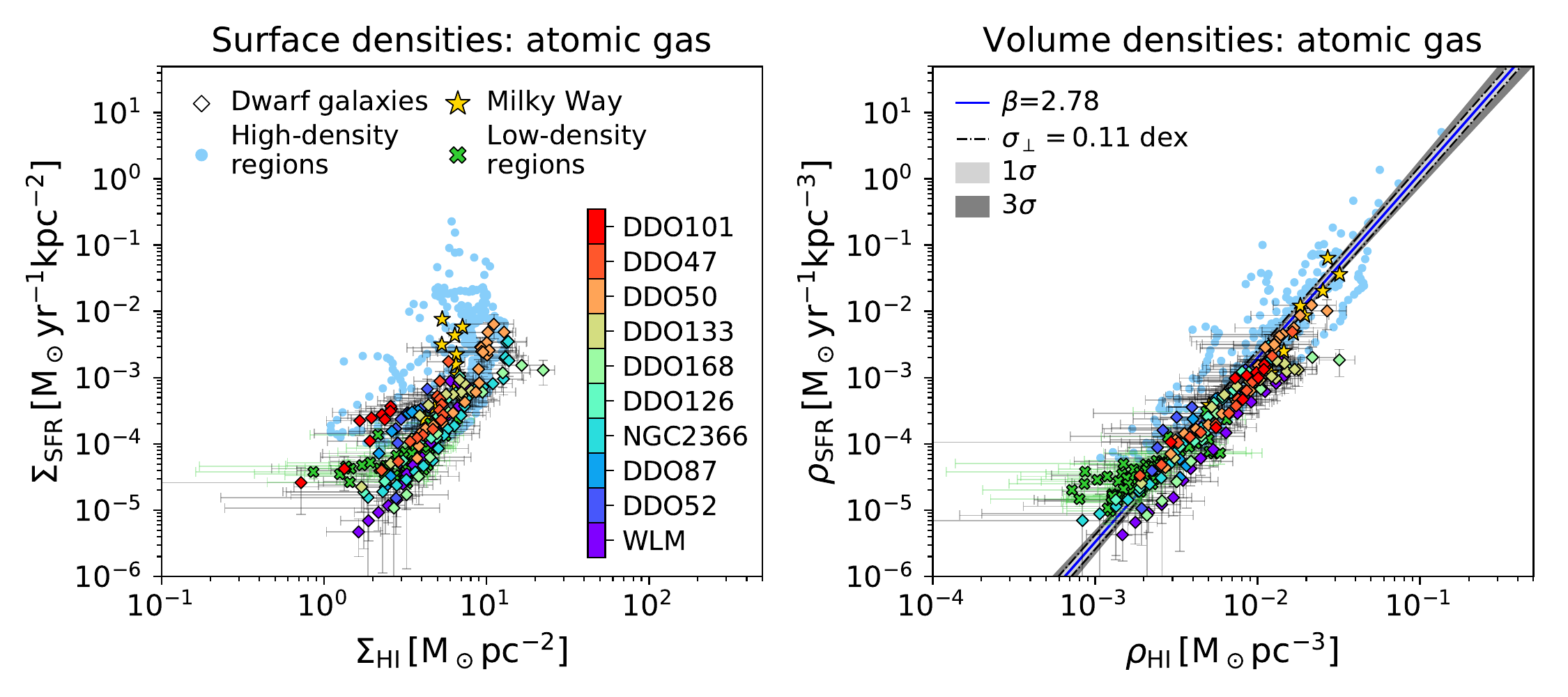}
\caption{Same as Fig.~\ref{fig:vsf}, but here the abscissae of the light blue points are the surface density (left) and the midplane volume density (right) of the atomic gas only, and the SFR scale height is the same as the atomic gas distribution (see text). 
The blue line shows the best-fit relation (Eq.~\ref{eq:log_vsf_HI}) obtained in this work.}
\label{fig:vsf_HI}
\end{figure*}

Since the HI is the dominant gas component of dwarf galaxies and a large part of the disc of spiral galaxies, investigating the link between the atomic gas and the SFR is of primary interest. 
The left panel of Fig.~\ref{fig:vsf_HI} shows that the new sample exhibits the same behaviour as both the low-density and the high-density regions of the more massive galaxies: there is no correlation between the atomic gas and the SFR surface densities or, if present, it is very weak and with a large scatter \citep[see also e.g.][]{1998Ferguson,2002Wong,2011Bolatto,2011Schruba,2019delosReyesKennicutt}. 

The right panel of Fig.~\ref{fig:vsf_HI} shows instead the correlation between the atomic gas and the SFR volume densities. 
The dwarf galaxies follow the same relation between the HI and the SFR volume densities found in \citetalias{2019Bacchini}
\begin{equation}\label{eq:log_vsf_HI}
 \log \left( \frac{\rho_\mathrm{SFR}}{\mathrm{M}_\odot \mathrm{yr}^{-1} \mathrm{kpc}^{-3}} \right) = 
 \beta
 \log \left( \frac{\rho_\mathrm{HI}}{\mathrm{M}_\odot \mathrm{pc}^{-3}} \right) 
 + \log B \, ,
\end{equation}
with $\beta = 2.79 \pm 0.08$, normalisation $\log B = 2.89_{-0.02}^{+0.03}$, and intrinsic scatter $\sigma_\perp = 0.12 \pm 0.01$~dex. 
We indeed repeated the fit of Eq.~\ref{eq:log_vsf_HI} including also the points of the dwarf galaxies and the low-density regions, finding that the parameters of the new best-fit are compatible within the errors with those reported in \citetalias{2019Bacchini} (see Table~\ref{tab:bestfit_vsf}). 
This extends the validity range of the HI-VSF law by about one order of magnitude. 

\renewcommand{\arraystretch}{1.5}
\begin{table}
        \centering
        \caption{Best-fit parameters of the VSF law with the total gas (Eq.~\ref{eq:log_vsf_gas}) and the relation with the atomic gas only (Eq.~\ref{eq:log_vsf_HI}). The first and second columns report the gas phases involved and the reference paper, respectively. The other columns provide the slope, the orthogonal intrinsic scatter $\sigma_\perp$, and the $y$-intercept with their uncertainties.}
        \label{tab:bestfit_vsf}
        \begin{tabular}{ll|ccc}
        	\hline\hline
        	Gas      & Ref.                      & Slope                  & $\sigma_\perp$           & $y$-intercept            \\
        	         &                           &                        & (dex)                    & (dex)           \\ \hline
        	HI+H$_2$ & \citetalias{2019Bacchini} & 1.91$^{+0.03}_{-0.03}$ & 0.12$^{+0.01}_{-0.01}$   & 0.90$^{+0.02}_{-0.02}$   \\
        	HI+H$_2$ & This work                 & 2.03$^{+0.03}_{-0.03}$ & 0.10$^{+0.01}_{-0.01}$   & 1.10$^{+0.01}_{-0.01}$   \\
        	HI       & \citetalias{2019Bacchini} & 2.79$^{+0.08}_{-0.08}$ & 0.13$^{+0.01}_{-0.01}$   & 2.89$^{+0.03}_{-0.02}$   \\
        	HI       & This work                 & 2.78$^{+0.06}_{-0.05}$ & 0.11$^{+0.01}_{-0.01}$   & 2.87$^{+0.02}_{-0.02}$ \\ \hline
        \end{tabular}
\end{table}

\section{Discussion}\label{sec:discussion}
Our new sample of dwarf galaxies follow the same relation as galaxies with higher masses and gas densities, suggesting the VSF law might be the general star formation law for nearby star-forming disc galaxies. 
In this section, we first quantify the impact of possible systematic effects on this result due to $i)$ the molecular gas content, $ii)$ the fraction of dust-obscured star formation, and $iii)$ the assumption of hydrostatic equilibrium. 
We also compare our work with other studies in the literature and, finally, we attempt some physical interpretation of our findings in order to obtain constraints on the mechanisms that may be relevant for star formation. 

\subsection{Systematic effects due to the molecular gas content}\label{sec:discussion_uncertainty_molgas} 
As mentioned earlier, CO emission lines in dwarf galaxies are faint or absent \citep[e.g.][]{1987Tacconi,2009Leroy,2011Bolatto,2012Schruba,2014Cormier,2015Hunt,2017Cormier}, but it is still unclear if this is due to the actual lack of molecular gas or simply to the low metallicity of this type of galaxies. 
Moreover, it is expected that some part of the molecular gas reservoir is not associated with CO in dwarf galaxies \citep[the so-called ``CO-dark'' gas; e.g.][]{2010Wolfire}. 
This is usually ascribed to a combination of efficient photo-dissociation of molecular clouds by the deeply penetrating UV radiation field from stellar clusters and the porosity of the ISM structure. 
As a consequence, CO is present only in small cores of molecular clouds, while it is photo-dissociated on larger spatial scales \citep[see e.g.][]{2015Cormier,2019MaddenCormier,2020Madden}. 
Instead, H$_2$ can efficiently self-shield from radiation, meaning that a significant amount of the molecular gas reservoir can exists outside CO-emitting regions. 
It is therefore interesting to investigate whether this molecular gas could significantly affect the VSF law shown in Fig.~\ref{fig:vsf}.

In our methodology, the fraction of molecular gas impacts both the gas and the SFR volume densities of dwarf galaxies. 
Given our definition of the molecular gas fraction as $f_\mathrm{H_2} = \Sigma_\mathrm{H_2} / \Sigma_\mathrm{gas}$, in Eq.~\ref{eq:rho_gas} we have therefore $\Sigma_\mathrm{H_2} = f_\mathrm{H_2} / (1-f_\mathrm{H_2}) \Sigma_\mathrm{HI}$ and, using the approximation $h_\mathrm{H_2} \approx h_\mathrm{HI} / 2$ (see \citetalias{2019Bacchini} and \citealt{2020Bacchini}), we obtain that the gas volume density increases as $\rho_\mathrm{gas} = (1+f_\mathrm{H_2})/(1-f_\mathrm{H_2}) \rho_\mathrm{HI}$. 
From Eq.~\ref{eq:hSFR}, we infer that the SFR scale height is $h_\mathrm{SFR} = h_\mathrm{HI} (2-f_\mathrm{H_2})/2$. 
Hence, the SFR volume density (Eq.~\ref{eq:rho_SFR}) increases of a factor of $2/(2-f_\mathrm{H_2})$ with respect to the case in which $f_\mathrm{H_2}=0$. 
In order to quantify the influence of $f_\mathrm{H_2}>0$ on our results, we explore the case with 30\% of molecular gas, which approximately corresponds to the estimates in the literature \citep[e.g.][]{2015Hunt,2019Hunter}
\footnote{The measurements of the molecular gas fraction in dwarf galaxies are uncertain, as $\alpha_\mathrm{CO}$ is poorly constrained. 
For example, the estimate obtained by \cite{2019Hunter} for LITTLE THINGS galaxies is based on the MW $\alpha_\mathrm{CO}$. 
Instead, \cite{2015Hunt}, who used a different sample of dwarf galaxies, assumed either the MW value or, for the most metal-poor galaxies, a scaling relation with metallicity, which is a power-law with index $\approx-2$. 
This index is also uncertain: for instance, \cite{2016Amorin} estimated $\approx -1.5$ and \cite{2019MaddenCormier} obtained $\approx -3.3$, which result in lower and higher molecular gas fractions, respectively.}. 
If $f_\mathrm{H_2}=0.3$, then $\rho_\mathrm{gas} \approx 1.9 \rho_\mathrm{HI}$ and $\rho_\mathrm{SFR}$ increases of $\approx 1.2$, producing a rightward shift of $\approx 0.28$~dex and an upward shift of $\approx 0.08$~dex on the points for dwarf galaxies in Fig.~\ref{fig:vsf}. 
The yellow pentagons in Fig.~\ref{fig:vsf_withcorrections} show the volume densities of our dwarf galaxies including $f_\mathrm{H_2}=0.3$. 
They lie systematically below the VSF law and the magenta diamonds obtained with $f_\mathrm{H_2}=0$, but this shift is comparable to the typical error-bar of our points, indicating that the effect of $f_\mathrm{H_2}$ is small and does not strongly impact the validity of the VSF law. 

\subsection{Systematic effects on the star formation rate surface density}\label{sec:discussion_uncertainty_obscuredSF}
\begin{figure}
\includegraphics[width=1.\columnwidth]{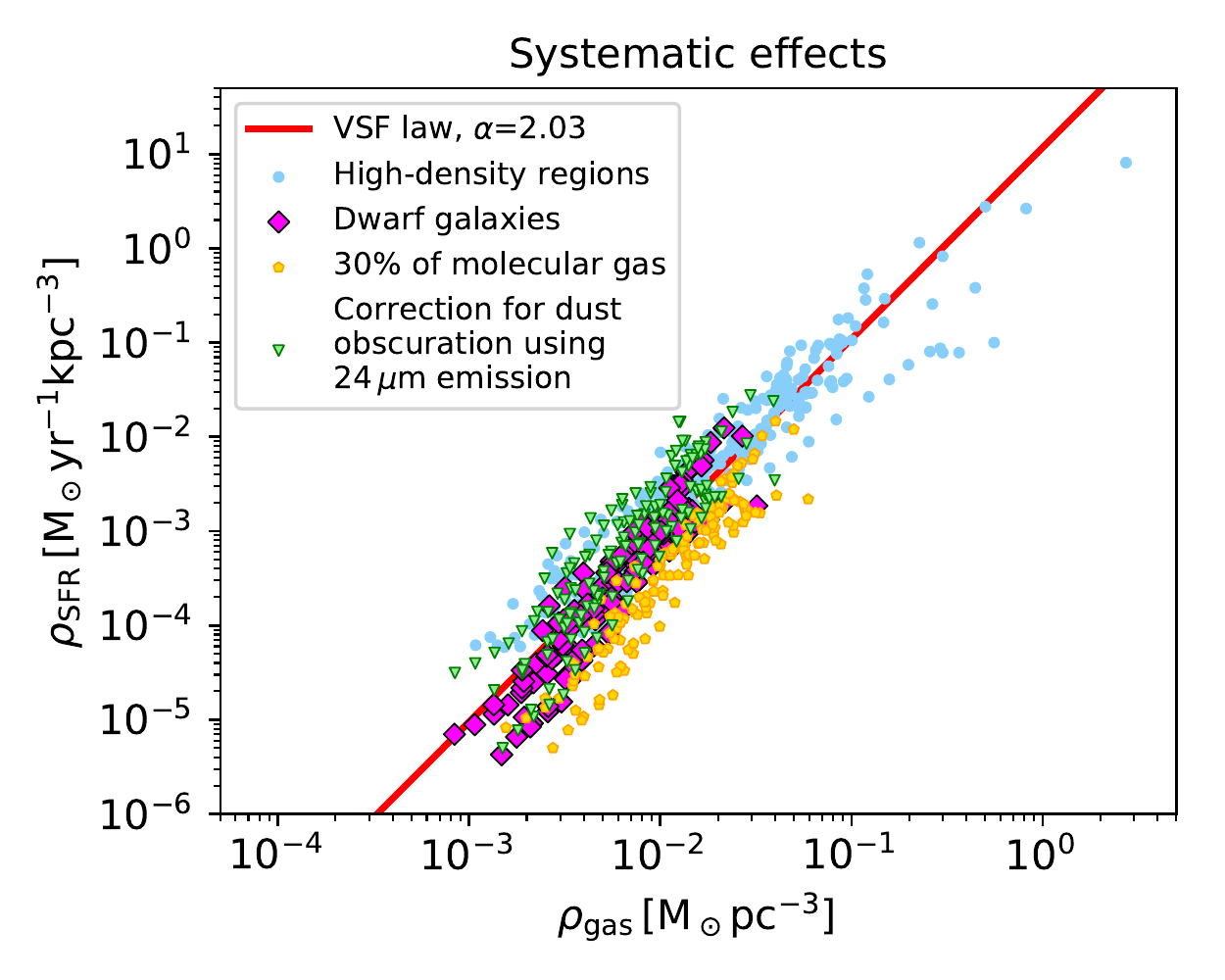}
\caption{Effect on the VSF law of including either a fraction of molecular gas of 30\% (yellow pentagons) or the correction for the dust obscuration using $24 \, \mu$m emission (green triangles). 
For comparison, the volume densities of our sample of dwarf galaxies from our fiducial analysis are shown by the magenta diamonds and are the same as in Fig.~\ref{fig:vsf}. 
The light blue points are from \citetalias{2019Bacchini} and the red line is the best-fit VSF law obtained in this work.}
\label{fig:vsf_withcorrections}
\end{figure}

The SFR surface densities adopted in \citetalias{2019Bacchini} included the correction for dust obscuration estimated from the $24 \, \mu$m luminosity \citep{2008Leroy}. 
Hence, in order to be fully consistent with this previous work, we should include this correction also for the new sample of dwarf galaxies. 
We recall however the amount of dust in dwarf galaxies is expected to be very low \citep[e.g.][]{2007Walter,2019MaddenCormier} and the $24 \, \mu$m luminosity is unlikely to have a strong impact on the SFR surface density. 
In order to quantify the effect of this component, we estimated the dust-corrected SFR through the relation reported by \cite{2008Leroy} and using the $24 \, \mu$m flux measured from \textit{Spitzer} observations by \cite{2009Dale} (except for DDO~47, which is not included in their sample) and the unobscured SFR using the FUV luminosity, which was re-scaled to include the reddening \citep{2012Zhang}. 
NGC~2366 and DDO~101 have the highest $24 \, \mu$m fluxes, while only upper limits are available for DDO~52 and DDO~87. 
The green triangles in Fig.~\ref{fig:vsf_withcorrections} show the effect of including the $24 \, \mu$m correction in the derivation of the SFR volume density. 
The agreement with the VSF law is conserved and the dwarf galaxies still follow the same trend as the galaxies studied in \citetalias{2019Bacchini}. 
On average, the $24 \, \mu$m-corrected SFR of our dwarf galaxies is increased of a factor $\approx 1.6$, corresponding to an upward shift of $\approx 0.2$~dex. 
We point out that the inclusion of the molecular gas fraction and the correction for dust obscuration tend to compensate, which indicates that our results would not change if both factors are taken into account. 

We note that the SFR of DDO~50 (a.k.a. Holmberg II) and NGC~2366 in Table~\ref{tab:sample} are a factor $\approx 2$ lower than the values 
obtained by \cite{2015McQuinn} using the CMD of resolved stellar populations. 
These authors reconstructed the galaxy star formation history by fitting the CMD with stellar evolutionary models and calculated the average SFR over the last 100~Myr, which corresponds to the mean age of FUV-emitting stars \citep{2012KennicuttEvans}. 
Aiming to assess the possible impact of this difference on our results, we re-scaled the SFR surface density radial profile adopted in this work in order to have the same SFR as \cite{2015McQuinn}. 
We found that using these modified profiles does not affect our main conclusion, as both DDO~50 and NGC~2366 remain in agreement with the VSF.

\subsection{Validity of the hydrostatic equilibrium}\label{sec:discussion_uncertainty_HE}
As discussed in Sect.~\ref{sec:results_scaleheight} and in \citetalias{2019Bacchini}, there is a general agreement between the scale heights based on the hydrostatic equilibrium derived by different authors, despite the different methods used to measure the velocity dispersion. 
We note however that the validity of the assumption of hydrostatic equilibrium for nearby galaxies has not been extensively tested in the literature. 
\citetalias{2019Bacchini_b} compared the radial profile of the scale height for our Galaxy derived assuming the hydrostatic equilibrium with that measured by \cite{2017Marasco} using emission-line observations, both for the atomic gas and the molecular gas distributions. 
It was found that, while the theoretical and the observed profiles are similar in the case of the molecular gas, the former tends to be $\approx 1.5$ times lower than the latter for the atomic gas. 
This discrepancy does not necessarily mean that the hydrostatic equilibrium is not valid, as it may also indicate the presence of a second HI component with high velocity dispersion (\citealt{2017Marasco}). 
Alternatively, one could guess that some anisotropic force (e.g. magnetic tension, cosmic rays) contributes to counteract the gravitational pull. 
However, \citetalias{2019Bacchini_b} found agreement between the (measured) scale heights of star formation tracers and the (modelled) scale height of the gas in hydrostatic equilibrium, suggesting that the gas which is converted into stars is indeed in hydrostatic equilibrium. 

Direct measurements of the gas scale height are of primary interest to test the validity of hydrostatic equilibrium and understand the structure of gaseous discs in galaxies. 
In the literature, there are some estimates of the gas scale height in edge-on galaxies \citep[e.g.][]{2011Yim,2014Yim,2017Peters,2020Yim}, but a comparison with the hydrostatic equilibrium is not provided. 
Moreover, highly inclined galaxies strongly suffer from projection effects: line-of-sight warps, non-circular motions and extra-planar gas can all artificially inflate measurements for both the velocity dispersion and the scale height of the gas \citep[e.g.][]{1997Swaters,1997Sicking,2007Oosterloo,2019Marasco_b}. 
Therefore, an ``ad hoc'' method applicable to galaxies with relatively high inclination is required in order to measure the scale height from the observations and avoid dramatic projection effects. 
This task is however beyond the scope of this work and we leave it to future investigations. 

It is worth to mention that the disc thickness itself can bias the galaxy properties obtained from line emission observations with respect to the intrinsic properties \citep{1997Sicking,2017Iorio,2018Iorio}. 
In particular, when the thickness is non-negligible, one tends to underestimate the intrinsic inclination of a galaxy, as its gas disc appears more face-on than it actually is. 
Consequently, the surface density profile and the rotation curve are underestimated and overestimated, respectively. 
\cite{2018Iorio} developed an innovative method to study the gas kinematics in thick discs based on combining the 3D tilted-ring modelling (i.e. $^{\mathrm{3D}}$\textsc{Barolo}) with the hydrostatic equilibrium (i.e. \textsc{Galpynamics}) using HI data cubes. 
This approach was tested on three dwarf galaxies, including WLM and NGC~2366, in order to derive their rotation curve, velocity dispersion, surface density, and scale height in a self-consistent way. 
\cite{2018Iorio} found that the thickness bias is of second order with respect to other possible sources of uncertainty (e.g. asymmetric-drift correction, inclination). 
The HI scale heights of WLM and NGC~2366 obtained by \cite{2018Iorio} are compatible with those shown in Fig.~\ref{fig:hHI_all}, indicating the bias due to the disc flaring is mild for our galaxies and does not significantly influence our results. 

\subsection{Comparison with other works}\label{sec:discussion_works}
It is interesting to compare the scale heights obtained with our method with those available in the literature and based on the assumption of vertical hydrostatic equilibrium. \cite{2011Banerjee} calculated the HI scale height for four dwarf galaxies, including DDO~50 (a.k.a. Holmberg II) and NGC~2366. 
They assumed that both the HI disc and the stellar disc are affected by their self-gravity. 
The only mass component of the external potential is the DM halo, which was assumed to have a pseudo-isothermal profile. 
We also note that, since these authors defined the scale height as the half width at half maximum (HWHM), their profiles should be divided by a factor of 1.177 when compared to ours. 
For DDO~50, these authors found $\text{HWHM }\approx 450$~pc over the whole disc, which is approximately compatible with our scale height for $R\gtrsim3$~kpc. 
However, our profile shows a clear flaring, despite the large uncertainties, while the scale height found by \cite{2011Banerjee} is practically constant. 
This discrepancy, which is mainly in the inner regions, can be explained by the differences in the velocity dispersion profile, which is slightly higher than ours for $R\lesssim3$~kpc, and in the mass model of the galaxy. 
For NGC~2366, \cite{2011Banerjee} found that the HWHM increases from $\approx 100$~pc in the inner regions to $\approx 800$~pc at $R\approx 5$~kpc assuming $\sigma_\mathrm{HI}$ constant at 9~\kms. 
This velocity dispersion is significantly different from the profile obtained by \cite{2017Iorio}, which explains the disagreement with our scale height for $R \lesssim 5$~kpc (see Fig.~\ref{fig:all_sdhi_vdhi_sfrd_1}).

In a recent study, \cite{2020Patra} derived the HI scale height for a sample of 23 dwarf galaxies from LITTLE THINGS, including also those in our sample. 
This author used the same method as \cite{2011Banerjee} to derive the scale height, which was defined as the HWHM of the distribution of the gas in hydrostatic equilibrium. 
In general, the HI scale height in this study shows a flaring for all the galaxies, which is in agreement with our results. 
For some galaxies in common with our sample, the HWHM calculated by \cite{2020Patra} is also compatible within the uncertainties with those in Fig.~\ref{fig:hHI_all} (e.g. DDO~50, DDO~87, DDO~133), but others are significantly different (e.g. WLM, DDO~47, DDO~101, DDO~126). 
For example, the flaring of the scale height is much steeper than ours in some disc outskirts (e.g. WLM) and the radial trend is clearly not monotonic with the galactocentric radius (e.g. DDO~47). 
These discrepancies might be partially due to the different mass models. 
Indeed, the HI rotation curves adopted by \cite{2020Patra} are significantly different from those used in our work (see \citealt{2017Iorio} for a discussion). 
However, the most important source of discrepancy in the flaring determination is likely the HI velocity dispersion (see Fig.~4 in \citealt{2020Patra} vs the central panels of Fig.~\ref{fig:wlm_sdhi_vdhi_sfrd} and Fig.~\ref{fig:all_sdhi_vdhi_sfrd_1}). 
\cite{2020Patra} derived the velocity dispersion by dividing the HI disc in rings and stacking, in each ring, the line profiles along the line of sight after shifting their centroid velocity to a common value in order to remove the contribution of rotation. 
This method can introduce an artificial broadening in the resulting profile if the stacked ones are not perfectly aligned, or if single profiles are described by multiple kinematic components (see \citealt{2017Iorio}). 
Instead, \cite{2017Iorio} used $^{\mathrm{3D}}$\textsc{Barolo}, which simultaneously fits (for each ring) the rotation velocity and the azimuthally averaged velocity dispersion in order to minimise the residuals between the data and the model. 
This markedly improves the reliability of velocity dispersion estimates with respect to other 2D methods (e.g. 2th moment map of the data cube, stacking or pixel-by-pixel fitting of the line profiles) also for data with low signal-to-noise ratio \citep{2015Diteodoro}.

Several authors have investigated the star formation law in different environments and interpreted their findings in the context of physical processes regulating star formation. 
For example, the SK law with $N \approx 1.4$ is usually explained by asserting that the timescale of the conversion of gas into stars is the free-fall time and assuming fixed scale heights for the gas and the SFR (i.e. $\Sigma_\mathrm{gas} \propto \rho_\mathrm{gas}$ and $\Sigma_\mathrm{SFR} \propto \rho_\mathrm{SFR}$). 
The break in the SK law at $\Sigma_\mathrm{gas} \sim 10 \mathrm{\, M_\odot pc^{-2}}$ is considered by some authors as an indication of a density threshold which needs to be exceeded to efficiently convert the gas into stars \citep[e.g.][]{2001MartinKennicutt,2004Schaye,2010Bigiel}, while other authors considered alternative star formation laws. 
In this respect, two possibilities are the so-called ``extended Schmidt law'' $\Sigma_\mathrm{SFR} \propto \Sigma_\mathrm{gas}^n \Sigma_\star^m$ \citep{1975TalbotArnett,2011Shi}, which is based on the idea that the existing stellar component with surface density $\Sigma_\star$ participate in regulating star formation, and the ``Silk-Elmegreen relation'' $\Sigma_\mathrm{SFR} \propto \Sigma_\mathrm{gas}/\tau_\mathrm{orb}$ \citep[e.g.][]{1997Silk,1997Elmegreen}, which sets the timescale of star formation to the galactic orbital time $\tau_\mathrm{orb}$. 

\cite{2018Shi} verified the extended Schmidt law for a large range of galactic environments, from the outermost star-forming regions of dwarf galaxies to spiral and merging galaxies, as well as individual molecular clouds in M33 \citep[see also][]{2011Shi,2017Roychowdhury}. 
They found that this relation has a scatter of about $0.3$~dex, which is larger than the $0.1$~dex of the VSF law. 
These authors also showed that both the SK law and the Silk-Elmegreen law are more scattered than the extended Schmidt law, with the outskirts of dwarf galaxies clearly deviating from these relations. 

Similarly, \cite{2019delosReyesKennicutt} found that dwarf galaxies tend to fall below the SK law, producing the break in the relation. 
However, this feature did not emerge clearly for the Silk-Elmegreen and extended Schmidt laws, which were found to be less scattered ($\sigma \approx 0.33$~dex) than the SK law including both spiral and dwarf galaxies ($\sigma \approx 0.37$~dex).

\cite{2008Leroy} compared the predictions of various models of star formation with the observed star formation efficiency (i.e. $\mathrm{SFE} = \Sigma_\mathrm{SFR} / \Sigma_\mathrm{gas}$) as a function of the galactocentric radius measured in dwarf and spiral galaxies. 
These authors investigated ``local'' star formation laws based on the idea that the timescale of the conversion of gas into stars is set by a given physical mechanism 
(i.e. gravitational instability, galactic rotation, cloud-cloud collisions, efficiency of molecular clouds in forming stars, and midplane gas pressure), but they did not find a single process that could describe both dwarf and spiral regimes. 
\cite{2008Leroy} also found that the observed SFE is not in agreement with the predictions of models based on a density threshold (i.e.~large-scale gravitational instability of the disc, molecular clouds destruction by shear, molecular gas formation), which distinguish whether the gas is dense enough to be star-forming. 
Hence, they concluded that the SFE is regulated by the interplay of multiple physical mechanisms acting on scales smaller than the spatial resolution of their data. 
The results found in this and the previous works \citepalias{2019Bacchini,2019Bacchini_b} indicate instead the existence of a unique local star formation law valid at all density regimes, which may suggest a simple and perhaps unique physical mechanism, involving exclusively the gas volume densities, at the core of star formation processes. 

Recently, \cite{2019Dey} performed a statistical analysis aiming to identify the strongest correlations between the SFR surface density and a set of stellar and molecular gas properties (e.g. surface density, velocity dispersion, metallicity) measured over kilo-parsec scale regions in a sample of 39 galaxies with stellar mass M$_\star \gtrsim 10^{10} \, M_\odot$. 
They found significant correlations involving the molecular gas and the stellar surface densities and estimated that including extra-factors in the star formation law based on the H$_2$ surface density reduces the scatter from about 0.3~dex to about 0.2~dex, still larger than the scatter of the VSF law.

Our approach is based on the assumption of vertical hydrostatic equilibrium, similarly to the analysis done by \cite{2006BlitzRosolowsky}. 
These authors concluded that star formation is regulated by the hydrostatic pressure of the gas, distinguishing between two regimes of star formation, one for low-pressure (HI-dominated) environments and the other for high-pressure (H$_2$-dominated) regions. 
These conclusions are somewhat different from ours as we do not find two regimes of star formation, but instead a monotonic relation valid for both the HI- and the H$_2$-dominated regions. 
This discrepancy is plausibly due to the different assumptions behind the equation of midplane pressure, as discussed in detail in \citetalias{2019Bacchini_b}. 
We recall that the most important differences between the approach of \cite{2006BlitzRosolowsky} and ours are that they neglected the dark matter distribution and, in addition, assumed that the gas velocity dispersion is constant with the galactocentric radius at 8~\kms\, for all the galaxies in their sample. 

The importance of the gas disc flaring in shaping the star formation law was also investigated by \cite{2015Elmegreen}, who developed a model which predicts that the index of the local SK law changes from $N=1.5$ to $N=2$ as a consequence of the gas being self-gravitating in the outskirts of spiral galaxies and in dwarf galaxies. 
Star formation is assumed to be regulated by gravity and the star formation law is written in terms of surface densities as $\Sigma_\mathrm{SFR} = \epsilon_\mathrm{ff} \Sigma_\mathrm{gas}/\tau_\mathrm{ff}$, where $\epsilon_\mathrm{ff}$ is the efficiency per unit free-fall timescale $\tau_\mathrm{ff}$ and it is assumed constant for all galaxies. 
In the main disc of spiral galaxies, the scale height was taken to be approximately constant with the galactocentric radius, hence $\rho_\mathrm{gas} \propto \Sigma_\mathrm{gas}$ and $\tau_\mathrm{ff} \propto \rho_\mathrm{gas}^{-1/2} \propto \Sigma_\mathrm{gas}^{-1/2}$. 
Therefore, the surface-based star formation law is $\Sigma_\mathrm{SFR} \propto \Sigma_\mathrm{gas}^{3/2}$, in agreement with the empirical SK law. 
In the outskirts of spiral galaxies and in dwarf galaxies, the gas disc was instead assumed self-gravitating, which implies that the gas scale height is $h_\mathrm{gas} = \sigma_\mathrm{gas}^2 / ( \pi G \Sigma_\mathrm{gas})$. 
This results in a different index for the surface-based star formation law, which is $\Sigma_\mathrm{SFR} \propto \Sigma_\mathrm{gas}^2 $ (assuming constant velocity dispersion). 
The observed break in the SK law is then explained in terms of projection effects due to the gas disc flaring rather than a consequence of a threshold density for star formation \citep{2018Elmegreen}, in agreement with our conclusions. 
We note however that there are a number of differences between this approach and ours, the most important being the model of the galactic potential. 
Indeed, we did not assume that the gas is distributed in a self-gravitating disc (neither in the main disc nor in the outskirts of galaxies), as we included the gravitational potential of the DM halo. 
This latter is the dominant mass component in dwarf galaxies and in the outskirts of spiral galaxies, hence it is very important to take into account the DM halo in these regimes in order to derive the gas scale height (see Fig.~3 in \citetalias{2019Bacchini}).

\subsection{On the physical implications of the VSF law}
We have found that the VSF is tighter than surface-based star formation laws and furthermore that it is valid for both dwarf and spiral galaxies, covering a range of volume densities from $\approx 7 \times 10^{-4}~\mathrm{M}_\odot \mathrm{pc}^{-3}$ to $\approx 2~\mathrm{M}_\odot \mathrm{pc}^{-3}$ for the gas and from $\approx 4 \times 10^{-5}~\mathrm{M}_\odot \mathrm{yr}^{-1} \mathrm{kpc}^{-3}$ to $\approx 10~\mathrm{M}_\odot \mathrm{yr}^{-1} \mathrm{kpc}^{-3}$ for the SFR. 
Within this range, the VSF law with the total gas is $ \rho_\mathrm{SFR} \propto \rho_\mathrm{gas}^\alpha$, where $\alpha \approx 2$. 
The lack of a break (i.e. change in the slope) has a crucial implication: there is no density threshold for star formation when the volume densities are considered (we cannot completely exclude though that some threshold may exist below $\rho_\mathrm{gas} \sim 10^{-3}~\mathrm{M_\odot pc^{-3}}$). 
The absence of a threshold implies that some star formation recipes implemented in large-scale numerical simulations and analytical models of galaxy evolution may need revision. 
The observed break in the SK law can thus be interpreted as due to the projection effects in the presence of disc flaring rather than to the drop of the SFE for densities below a certain threshold (see also \citealt{1974Madore,1998Ferguson}; \citetalias{2019Bacchini,2019Bacchini_b}; but see \citealt{2020Kumari} for a different perspective). 
This was also suggested by \cite{2018Elmegreen} for the star formation law based on the surface densities using a model in which star formation is controlled by gravity (see Sect.~\ref{sec:discussion_works}). 

It is worth to discuss the possible role of azimuthal averages for the absence of a density threshold. 
The VSF law was derived using azimuthally averaged radial profiles, while the presence of a threshold for star formation is sometimes found when the surface densities are measured locally, by a pixel-by-pixel analysis for instance \citep[e.g.][]{2008Bigiel}. 
\cite{2013Boissier_book} showed indeed that the break in the SK law involving azimuthal averages can be ``smoothed'' if the gas disc is characterised by super-critical (star-forming) and sub-critical (non star-forming) regions, like in the presence of spiral arms for example. 
Depending on the spiral arms covering fraction, a galaxy would then follow a star formation law with different slope and ``apparent'' threshold density. 
Despite the samples considered in our work include both galaxies with and without spiral arms, we find neither indications of a change in the slope of the VSF law nor that the galaxies with spiral arms follow a different relation than those without. 
Moreover, if each galaxy follows a relation with different slope and density threshold than the others, the scatter between the points will increase in the low-density regime. 
This feature should be present in the surface-based as well as in the volume-based relations, as our method cannot correct for this effect. 
Instead, the scatter is significantly reduced in the right panel of Fig.~\ref{fig:vsf} with respect to the left panel, indicating that the main source of scatter in the surface-based relation is indeed the projection effect due to the disc flaring.

We recall that our method is suitable to derive the volume densities in the galactic midplane (Eqs.~\ref{eq:rho_gas} and~\ref{eq:rho_SFR}). 
Despite that, the midplane density is, for any given vertical distribution, proportional to the average density at a certain location in the disc, hence the validity of the VSF law may not be limited to the midplane. 
In the light of this, our results may suggest that the local average SFR volume density in a galaxy disc is regulated by the local average volume density of the gas. 
In order to gain insight on the mechanisms regulating the conversion of gas into stars, let us write the ``theoretical'' VSF law as
\begin{equation}\label{eq:interp_rho}
 \rho_\mathrm{SFR} = \epsilon \frac{\rho_\mathrm{gas}}{\tau_\mathrm{sf}} \, ,
\end{equation}
where $\epsilon$ is the efficiency per unit star formation timescale $\tau_\mathrm{sf}$. 
Given that the volume densities obtained through Eq.~\ref{eq:rho_gas} and Eq.~\ref{eq:rho_SFR} are those at $z=0$, we can speculate that they are a good approximation of the (azimuthally averaged) volume density close to the midplane of a galaxy, where most of the star-forming gas is concentrated. 
We must note though that it is not clear to what extent empirical star formation laws can capture the complexity of the conversion of gas into stars, hence it is important to bear in mind that the VSF law is meaningful on kiloparsec scale but likely not applicable to clouds or filaments. 
If we assume that $\tau_\mathrm{sf}$ is equal to the free-fall timescale of the gravitational instability $\tau_\mathrm{ff} \propto (G \rho_\mathrm{gas})^{-1/2}$ \citep[e.g.][]{1977Madore}, it follows from Eq.~\ref{eq:interp_rho} that the index of the ``empirical'' VSF law (i.e. $\alpha \approx 2$) can be obtained only with $\epsilon \propto \rho_\mathrm{gas}^{1/2}$. 
The origin of this proportionality is not clear and may arise from the combination of different factors, such as a radial variation of the molecular gas fraction \citep[e.g.][]{2015ElmegreenHunter}, the metallicity gradient, or the degree of ionisation of the gas \citep[e.g.][]{2007MccKeeOstriker}. 
These possibilities might be tested by looking for correlations between the scatter of the VSF law and the properties of our galaxies. 
The intrinsic scatter of our relation is however very small (0.1~dex), suggesting that secondary correlations are absent or too weak to be revealed.

Alternatively, we can think of some physical mechanism which may play a role in star formation and whose timescale is $\tau_\mathrm{sf} \propto \rho_\mathrm{gas}^{-1}$ (assuming constant efficiency). 
In order to fragment in gravitationally-bound (potentially star-forming) clouds, the diffuse ISM must first lose significant part of its thermal energy through radiative cooling. 
This suggests that, while the free-fall time likely governs star formation at the scales of molecular clouds, on larger (kiloparsec) scales the cooling time may be equally and perhaps more important than the free-fall time.
Similar ideas were also proposed by other authors, who suggested that $\tau_\mathrm{sf}$ might be the timescale of the slowest and ``bottle-neck'' process among collapse, cooling, and molecule formation \citep[e.g.][]{2007CiottiOstriker,2013Krumholz}. 
Both the cooling time and the timescale to reach the equilibrium between formation and destruction of H$_2$ are inversely proportional to the gas density \citep[e.g.][]{1979Hollenbach,2014Krumholz_rev}, resulting in $\alpha=2$ in Eq.~\ref{eq:interp_rho} and thus in agreement with the empirical VSF law. 

Finally, let us briefly discuss the possible origin of the volumetric relation with the atomic gas only, which is quite surprising. 
This correlation appears to have a larger scatter in the high-density regions of galaxies, where the molecular gas is detected using CO emission, than in the other parts of the discs (see Fig.~7a in \citetalias{2019Bacchini}). 
This may indicate that, in the inner and high-density regions, the total gas is a better tracer of the star-forming gas than the HI only, but this latter becomes a good tracer where the gas density is lower \citep[see also][]{2015ElmegreenHunter,2016Hu}. 
Possibly, this can be explained by the presence of a significant amount of CO-dark gas in the outskirts of galaxies. 
Alternatively, the relation involving the atomic gas only might be considered an indication that star formation can directly occur in the atomic gas in particular conditions (see e.g. \citealt{2012GloverClark,2012KrumholzHI}). 
We note that this correlation might open up to semi-empirical studies of galaxy evolution in low-density and metal-poor regions, where the molecular gas emission is not detected or very uncertain and some theoretical models of star formation yields discordant conclusions (see e.g. \citealt{2010Ostriker,2013Krumholz}; \S~4 and \S~6 in \citealt{2017Knapen_book} and references therein).

\section{Summary and conclusions}\label{sec:conclusions}
The star formation law is a key relation to link the gas content of a galaxy and its SFR, fundamental to understand galaxy formation and evolution. 
However, when we observe the gas and the SFR distributions in a galaxy, we can directly measure only their projected surface densities, while the intrinsic volume densities, which arguably are more physically meaningful, remain inaccessible. 
The flaring of gas discs in galaxies complicates the task of reconstructing intrinsic volume densities from observed surface densities, preventing to derive the intrinsic distributions in a straightforward way. 

\citetalias{2019Bacchini} and \citetalias{2019Bacchini_b} have developed and consolidated a method to convert the observed surface density radial profiles of the gas and the SFR to the corresponding volume density profiles using the scale height of their vertical distribution. 
This approach is based on the assumption of hydrostatic equilibrium and requires the knowledge of the gravitational potential of a galaxy and the velocity dispersion of the gas. 
Using this method, the volume densities of the gas and the SFR were derived for a sample of 12 nearby galaxies and for the MW. 
These two quantities were found to tightly follow one single power-law relation, the volumetric star formation (or VSF) law, which has a smaller scatter than the surface-based star formation laws. 
An unexpected correlation between the volume densities of the atomic gas and the SFR also emerged from these studies. 

The main aim of the present work was to extend the VSF law to the regime of dwarf galaxies, which is of primary importance to investigate the presence of a density threshold for star formation in low-density and HI-dominated environments. 
As a consequence of the shallow gravitational potential, the gas discs in this type of galaxies are thick and significantly flaring, hence taking into account the projection effects is fundamental. 
We applied the method used in \citetalias{2019Bacchini} to a sample of ten dwarf galaxies with robust HI kinematics and mass models available in the literature \citep{2017Iorio,2017Read}. 
The outermost star-forming regions (i.e. beyond the stellar disc) of the galaxies studied in \citetalias{2019Bacchini} were also added in this work, as they are low-density and HI-dominated parts of the disc where the flaring is prominent. 
We verified that both the new sample of dwarf galaxies and the star-forming outskirts follow the VSF law, extending its validity range down to $\rho_\mathrm{HI} \sim 10^{-3}~\mathrm{M_\odot pc^{-3}}$ and $\rho_\mathrm{SFR} \sim 10^{-5}~\mathrm{M_\odot yr^{-1} kpc^{-3}}$. 
We confirm both the VSF law with the total gas (i.e. HI+H$_2$) and the correlation involving the atomic gas only. 
Hence, the conclusions of this work are the following.
\begin{enumerate}[\textbullet]
\item The VSF law, namely $\rho_\mathrm{SFR} \propto \rho_\mathrm{gas}^\alpha$ with $\alpha \approx 2$, is valid for both low-density and high-density star-forming environments in nearby disc galaxies. 
The intrinsic scatter of the VSF law is $\sigma_\perp \approx 0.1$~dex, which is significantly lower than that of the star formation laws based on the surface densities. 
\item The atomic gas volume density correlates with the SFR volume densities, following the relation $\rho_\mathrm{SFR} \propto \rho_\mathrm{HI}^\beta$ with $\beta \approx 2.8$ and intrinsic scatter $\sigma_\perp \approx 0.1$~dex. 
This indicates that, contrary to previous claims based on projected surface densities, the atomic gas is a reliable tracer of the star-forming gas, which can be particularly useful in the low-metallicity regions of galaxies where the emission of molecular gas tracers is not detected. 
\item We find no evidence for a break in the VSF law occurring at the transition to the low-density and HI-dominated parts of the discs, which disfavours the existence of a density threshold for star formation.
\end{enumerate}

It is remarkable that the VSF law is valid for both dwarf and spiral galaxies, and for such a wide range of densities. 
This may indicate that our relation is the fundamental star formation law, also considering that volume densities are intrinsic quantities that are likely more physically meaningful than surface densities, which depend on projection effects. 
The VSF law is consistent with the idea that, on kiloparsec scale, gas cooling might be the primary process which sets the conversion of gas into stars, beforehand the gravitational instability is at play on cloud scales. 
Future investigations could aim to test the VSF law in environments that are even more extreme than dwarf galaxies, such as early-type galaxies with ongoing star formation, extended star-forming discs \citep[e.g.][]{2007Thilkera,2007Thilkerb}, and starbursts.

\section*{Acknowledgements}
We thank Bruce Elmegreen for the stimulating discussions and for providing thorough comments on this manuscript, and Deidre Hunter for sharing the surface photometry data and the support in their usage. 
CB would like to thank Giuliano Iorio for the help with \textsc{Galpynamics}. 
GP acknowledges support by the Swiss National Science Foundation, grant PP00P2\_163824.
%

\bibliographystyle{aa}
\bibliography{paty.bib}

%
\onecolumn
\begin{appendix}
	
\section{Radial profiles of $\Sigma_\mathrm{HI}$, $\sigma_\mathrm{HI}$, and $\Sigma_\mathrm{SFR}$}\label{ap:obs_ALL}
In Fig.~\ref{fig:all_sdhi_vdhi_sfrd_1}, we show the radial profiles of the atomic gas surface density and velocity dispersion and the SFR surface density used in this work for nine out of ten galaxies in our sample (see Fig.~\ref{fig:wlm_sdhi_vdhi_sfrd} for the profiles of WLM). 
The atomic gas profiles are from \cite{2017Iorio}, while the SFR surface density is obtained from the FUV photometry \citep{2012Zhang}.
\begin{figure*}[hp]
\includegraphics[width=1\columnwidth]{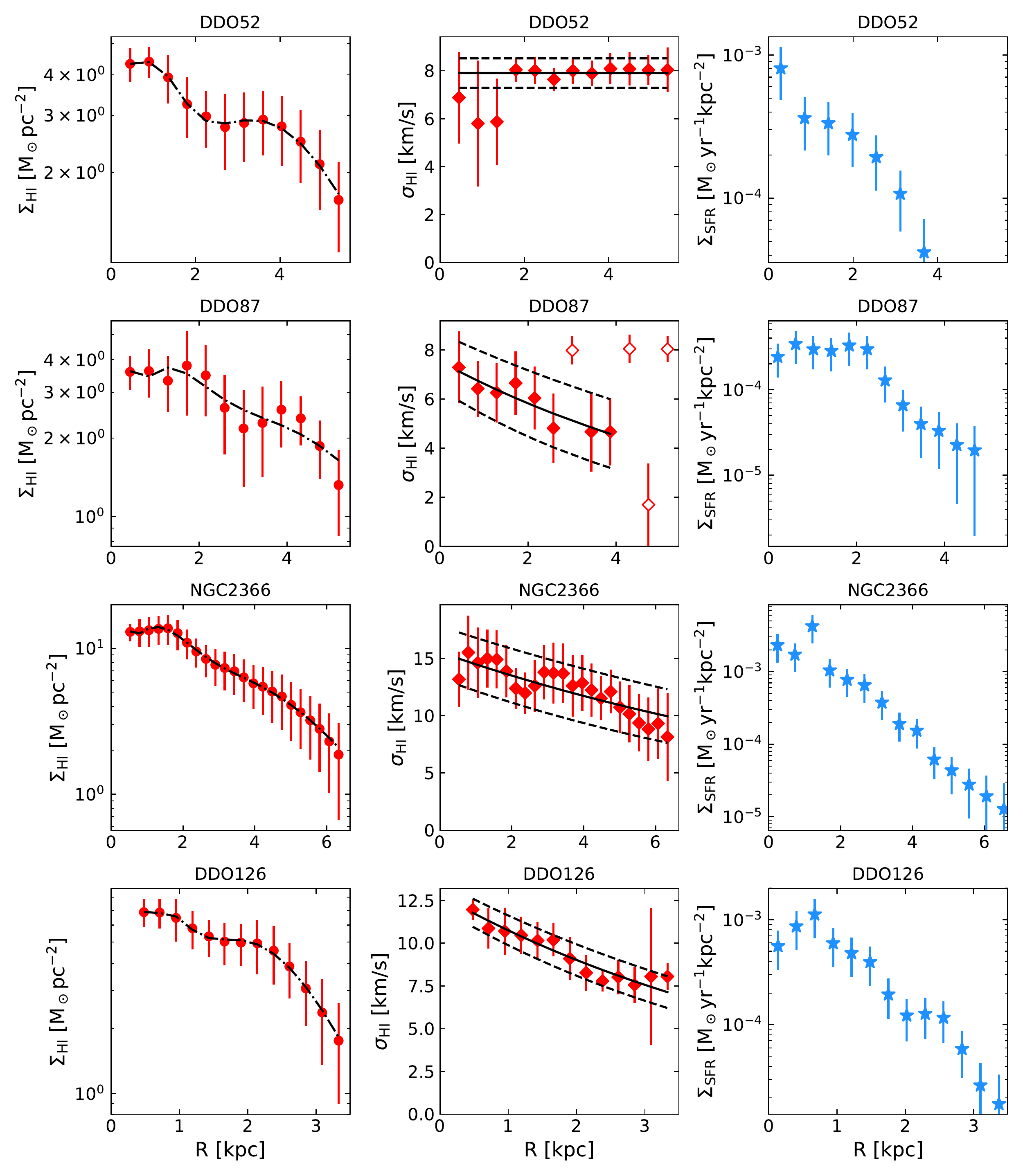}
\caption{Azimuthally averaged radial profiles of the atomic gas surface density (left), the HI velocity dispersion (centre), and the SFR surface density (right) of the dwarf galaxies studied in this work. 
The black curves are the same fits as in Fig.~\ref{fig:wlm_sdhi_vdhi_sfrd}.}
\label{fig:all_sdhi_vdhi_sfrd_1}
\end{figure*}
\begin{figure*}
\ContinuedFloat
\includegraphics[width=1.\columnwidth]{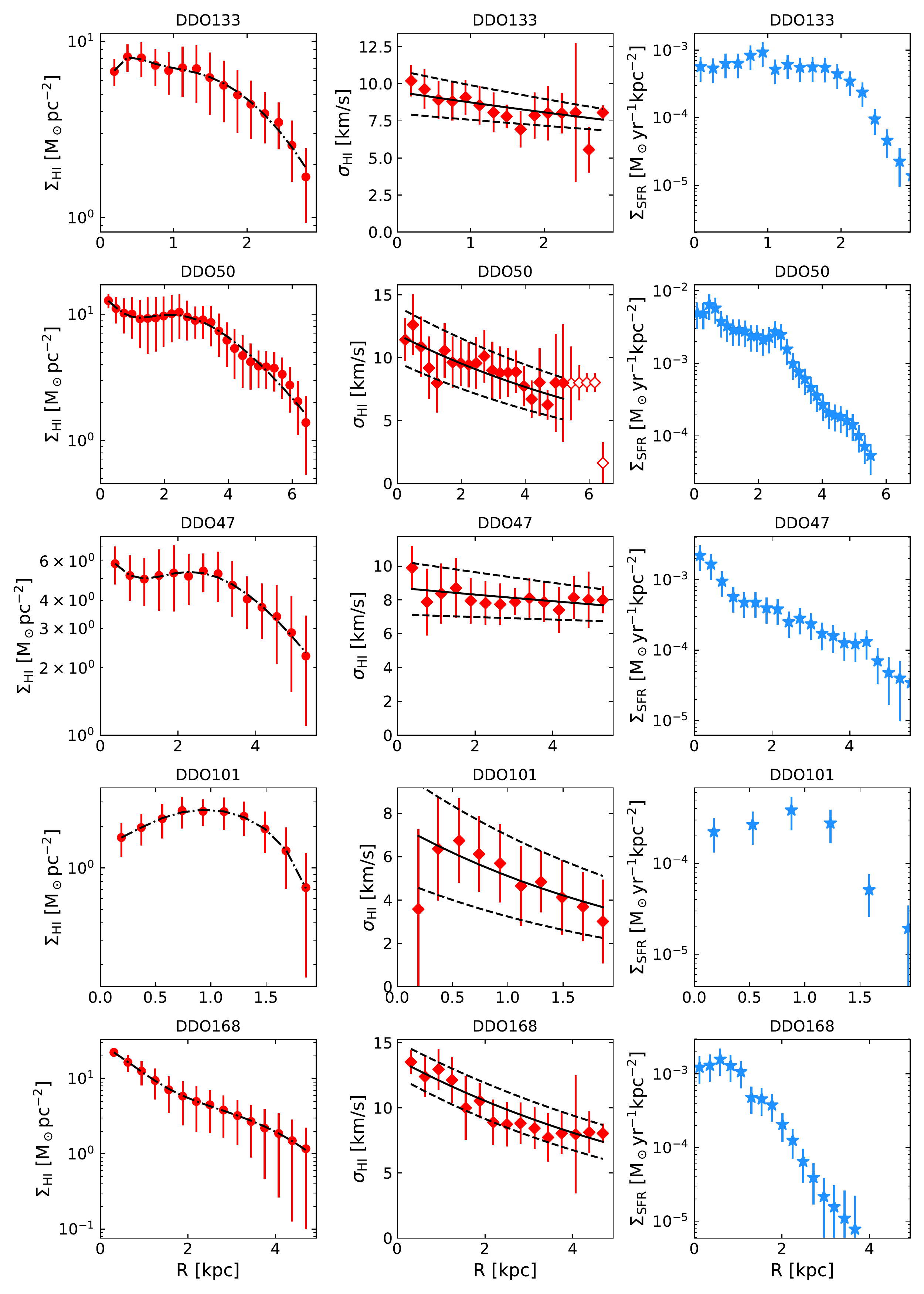}
\caption{Continued.}
\end{figure*}

\end{appendix}

\end{document}